\documentclass[10pt,twocolumn,twoside]{IEEEtran}
\renewcommand{\baselinestretch}{0.95}

\usepackage{color}
\usepackage{theorem}
\usepackage{times,amsmath,epsfig}
\usepackage{amssymb}
\usepackage{cite}

\usepackage{algorithmic}
\usepackage{algorithm}
\usepackage[font=small,format=plain,up,up]{caption}
\usepackage{subcaption}
\usepackage{needspace}

\usepackage{tikz}
\usetikzlibrary{shapes,arrows}
\input{mysymbol.sty}
\tikzstyle{phantom vertex} = [ ellipse, 
                               anchor = center, 
                               minimum height = 1*\unit, 
                               minimum width  = 1*\unit,
                               inner sep=0pt,
                               anchor=center]
\tikzstyle{red vertex}   = [black, fill = red!20,   phantom vertex, draw]
\tikzstyle{black vertex} = [black, fill = black!20, phantom vertex, draw]
\tikzstyle{blue vertex}  = [black, fill = blue!20,  phantom vertex, draw]
\tikzstyle{green vertex} = [black, fill = green!20,  phantom vertex, draw]
\tikzstyle{yellow vertex} = [black, fill = yellow!20,  phantom vertex, draw]
\tikzstyle{cyan vertex} = [black, fill = cyan!20,  phantom vertex, draw]
\tikzstyle{white vertex} = [black, fill = white!20,  phantom vertex, draw]
\tikzstyle{vertex}       = [draw, phantom vertex]

\tikzstyle{point} = [ellipse, inner sep=0pt, draw, fill=white, anchor = center,
                     minimum height = 0.05*\unit, minimum width  = 0.05*\unit]

\newcommand{\mymod}{\mathrm{mod}_N}
\newcommand{\trace}{{\mathrm{trace}}}
\newcommand{\bbeps}{{\boldsymbol{\epsilon}}}

\newtheorem{mylemma}{\bf Lemma}
\newtheorem{myproposition}{\bf Proposition}
\newtheorem{remark}{\bf Remark}

\psfull

\title{Reconstruction of Graph Signals through Percolation from Seeding Nodes}

\author{\IEEEauthorblockN{Santiago Segarra, Antonio G. Marques, Geert Leus, and Alejandro Ribeiro}
\thanks{Work in this paper is supported by Spanish MINECO grant No TEC2013-41604-R and USA NSF CCF-1217963. S. Segarra and A. Ribeiro are with the Dept. of Electrical and Systems Eng., Univ. of Pennsylvania.  A. G. Marques is with the Dept. of Signal Theory and Comms., King Juan Carlos Univ. G. Leus is with the Dept. of Electrical Eng., Mathematics
and Computer Science, Delft Univ. of Technology. Emails: ssegarra@seas.upenn.edu, antonio.garcia.marques@urjc.es, g.j.t.leus@tudelft.nl, and aribeiro@seas.upenn.edu. Parts of this paper will be presented at EUSIPCO 2015 \cite{EUSIPCO_our_interp_2015} and GLOBALSIP 2015 \cite{GlobalSIP_our_interpolation_2015}.}}

\begin{document}
\maketitle


\begin{abstract}%
New schemes to recover signals defined in the nodes of a graph are proposed. Our focus is on reconstructing bandlimited graph signals, which are signals that admit a sparse representation in a frequency domain related to the structure of the graph.
Most existing formulations focus on estimating an \emph{unknown} graph signal by \emph{observing} its value on a subset of nodes. By contrast, in this paper, we study the problem of reconstructing a \emph{known} graph signal using as input a graph signal that is non-zero only for a small subset of nodes (seeding nodes). The sparse signal is then percolated (interpolated) across the graph using a graph filter. Graph filters are a generalization of classical time-invariant systems and represent linear transformations that can be implemented distributedly across the nodes of the graph. Three setups are investigated. In the first one, a single simultaneous injection takes place on several nodes in the graph. In the second one, successive value injections take place on a single node. The third one is a generalization where multiple nodes inject multiple signal values. For noiseless settings, conditions under which perfect reconstruction is feasible are given, and the corresponding schemes to recover the desired signal are specified. Scenarios leading to imperfect reconstruction, either due to insufficient or noisy signal value injections, are also analyzed. Moreover, connections with classical interpolation in the time domain are discussed. The last part of the paper presents numerical experiments that illustrate the results developed through synthetic graph signals and two real-world signal reconstruction problems: influencing opinions in a social network and inducing a desired brain state in humans.
\end{abstract}


\section{Introduction}\label{S:Introduction}

Sampling and reconstruction of bandlimited signals are cornerstone problems in classical signal processing. The emergence of new fields of knowledge such as network science and big data is generating a pressing need to extend the results existing for classical time-varying signals to signals defined on graphs \cite{EmergingFieldGSP,SandryMouraSPG_TSP13,RabICASSP12_ApproxSignalsGraphs}. This not only entails modifying the existing algorithms, but also gaining intuition on the concepts that are preserved and lost when a signal is defined not in the time grid, but in a more general graph domain. In the context of reconstruction of graph signals, two different approaches have been developed. A first approach is related to the interpolation of bandlimited signals, {which consists in} inferring unobserved values by leveraging the fact that the signal lives in a low-dimensional space \cite{OrtegaInterpolation_icassp13,SamplingKovacevic_without_Moura_15, Oursampling_journal_2015}. Although most interpolation approaches are centralized, iterative \cite{OrtegaInterpolation} and distributed \cite{wang2015local, wang2015DistributedTrackGraphSig} interpolation schemes have also been developed. A different approach towards graph signal reconstruction is graph regularization \cite{SamplingKovacevicMoura_1415, zhou2004regularization} where a notion of smoothness is assumed on the signal and the unobserved values are estimated based on this notion. Both approaches coincide in that they estimate a graph signal from the observation of a subset of the signal values. {By contrast, the approach in this paper is to preserve the two-step methodology used when recovering bandlimited time-varying signals, which consists in the generation of a sparse signal followed by the application of a low-pass filter to reconstruct the missing entries, and extend it to the more general graph domain.}

{To be more specific, we} study the reconstruction of bandlimited graph signals through the application of {low-pass} \emph{graph filters} to sparse \emph{seeding signals}. Graph filters are the generalization of classical time-invariant systems when the signals are defined on a general graph as opposed to the classical time domain \cite{SandryMouraSPG_TSP13}. Seeding signals are graph signals attaining nonzero values on a subset of the nodes in the graph, called seeding nodes.
To describe our approach more precisely, let $\bby$ stand for the target graph signal we want to recover. Our goal is to design a graph filter $\bbH$ and a sparse signal $\bbx$ such that $\bby$ can be obtained upon applying $\bbH$ to $\bbx$. The design is accomplished in two steps. In the first step, we design the filter $\bbH$ leveraging the bandlimitedness of $\bby$, to eliminate the frequencies not present in $\bby$. Then, we use the $\bbH$ designed in the first step and the specific value of $\bby$ to design the signal $\bbx$. The challenge is that $\bbx$ cannot be chosen freely but is rather the output of a \emph{seeding phase} where only a few seeding nodes inject values. {The seeding phase requires more elaboration than its counterpart for time-varying signals, not only because graph signals are less regular, but also because it will be shown that in general the seeding values cannot coincide with those of the signal to recover.} For a rigorous problem definition see Section \ref{S:signal_reconstruction_graph_filters} {and Figure~\ref{F:example_seedings} in Section \ref{S:Extrapolating_GS_subset_nodes}.} Since graph filters act on graph signals through the successive application of local operators, the output of a graph filter can be viewed as the outcome of a diffusion or \emph{percolation} process. { Applications include the generation of an opinion profile in a social network \cite{WattsDodds07} by influencing a few agents (Section~\ref{sec_num_social_networks}) and the synthesis of brain signals \cite{hallett2000transcranial} by exciting a few neural regions (Section~\ref{sec_num_brain_state}). Other potential applications for signal reconstruction via local interactions include molecular communications in nanonetworks \cite{Nakano_etal12, Kuran201086} and wireless sensor networks \cite{Intanagonwiwat_etal00}.}

The paper investigates three different reconstruction schemes, each of them associated with a different seeding phase. In Section \ref{S:Extrapolating_GS_subset_nodes}, the seeding phase consists of a unique seeding signal with several nonzero values, which coincides with the intermediate signal $\bbx$. By contrast, in Section \ref{S:Extrapolating_GS_succesive} the seeding phase consists of several seeding signals injected by a single seeding node. At each instant, the signal is percolated (diffused) within \textit{one-hop} neighborhoods. The support of $\bbx$ depends on the duration of the seeding phase and the connectivity of the seeding node. Finally, in Section \ref{S:shift_space_extrapolation} we consider a more general scheme which merges the two earlier approaches. In this scheme, the seeding phase consists of several time instants and, in each of them, multiple nodes are allowed to inject a signal. The schemes will be referred to as multiple node-single time (MN-ST), single node-multiple time (SN-MT) and multiple node-multiple time (MN-MT) seeding, respectively. For the three of them, we state conditions on the underlying graph and the seeding nodes to guarantee perfect reconstruction of any bandlimited signal.
We also show that, in general, if the interpolator takes the form of a graph filter, the seeding values cannot coincide with those of the signal to interpolate. Furthermore, we discuss how additional seeding values can be used to reduce the complexity of the graph filter needed for perfect recovery and draw connections with classical interpolation of time-varying signals. In Section \ref{sec_imperfect_reconstruction} we study the reconstruction performance in imperfect settings, either because the seeding values are insufficient in number or corrupted by noise. In Section \ref{S:NumExper} we run numerical experiments to illustrate signal reconstruction in noiseless and noisy scenarios using both synthetic and real-world graphs.\footnote{\textbf{Notation:} $\mathbf{e}_i$ is the $i$th $N\times 1$ canonical basis vector (all entries of $\mathbf{e}_i$ are zero except the $i$th one, which is one); $\mathbf{E}_K:=[\mathbf{e}_1,...,\mathbf{e}_K]$ is a tall matrix collecting the $K$ first canonical basis vectors while $\bar{\mathbf{E}}_K:=[\mathbf{e}_{K+1},...,\mathbf{e}_N]$ collects the last $N-K$ canonical basis vectors; $\mathbf{0}$ and $\mathbf{1}$ are, respectively, the all-zero and all-one matrices (when not clear from the context, a subscript indicating the dimensions will be used).}

\section{Bandlimited graph signals and graph filters}\label{S:Modeling}

Let $\mathcal{G}$ denote a directed graph with a set of $N$ nodes or vertices $\mathcal{N}$ and a set of links $\mathcal{E}$, such that if node $i$ is connected to $j$, then $(i,j)\in\mathcal{E}$. The (incoming) neighborhood of $i$ is defined as the set of nodes $\mathcal{N}_i = \{j \,| \, (j,i)\in\mathcal{E}\}$ connected to $i$. For any given graph we define the adjacency matrix $\bbA$ as a sparse $N\times N$ matrix with nonzero elements $A_{ji}$ if and only if $(i,j)\in\ccalE$. The value of $A_{ji}$ captures the strength of the connection from $i$ to $j$. The focus of this paper is not on analyzing $\mathcal{G}$,  but a graph signal defined on the set of nodes $\mathcal{N}$. Formally, such a signal can be represented as a vector $\bbx=[x_1,...,x_N]^T \in  \mathbb{R}^N$ where the $i$-th component represents the value of the signal at node $i$ or, alternatively, as a function $f : \mathcal{N} \to \mathbb{R}$, defined on the vertices of the graph.

The graph $\mathcal{G}$ is endowed with a \emph{graph-shift operator} $\bbS$ \cite{SandryMouraSPG_TSP13,SandryMouraSPG_TSP14Freq}. The shift $\bbS$ is a $N\times N$ matrix whose entry $S_{ji}$ can be nonzero only if $i=j$ or if $(i,j)\in\mathcal{E}$. The sparsity pattern of the matrix $\bbS$ captures the local structure of $\ccalG$, but we make no specific assumptions on the values of the nonzero entries of $\bbS$. Choices for $\bbS$ are the adjacency matrix of the graph \cite{SandryMouraSPG_TSP13,SandryMouraSPG_TSP14Freq}, its Laplacian \cite{EmergingFieldGSP}, and their respective generalizations \cite{godsil2001algebraic}. The intuition behind $\bbS$ is to represent a linear transformation that can be computed locally at the nodes of the graph. More rigorously, if $\bby$ is defined as $\bby=\bbS\bbx$, then node $i$ can compute $y_i$ provided that it has access to the value of $x_j$ at $j\in \mathcal{N}_i$. We assume henceforth that $\bbS$ is diagonalizable, so that there exists a $N\times N$ matrix $\bbV$ and a $N\times N$ diagonal matrix $\bbLambda$ that can be used to decompose $\bbS$ as $\bbS=\bbV\bbLambda\bbV^{-1}$.
In particular, $\bbS$ is diagonalizable when it is normal, i.e., it satisfies $\bbS\bbS^H=\bbS^H\bbS$ where $\bbS^H$ denotes the conjugate transpose of $\bbS$. In that case, we have that $\bbV$ is unitary, which implies $\bbV^{-1}=\bbV^{H}$, and leads to the decomposition $\bbS=\bbV\bbLambda\bbV^H$.

We are interested in cases where the graph-shift operator $\bbS$ plays a role in explaining the graph signal $\bbx$. More specifically, cases where $\bbx$ can be expressed as a linear combination of a \emph{subset} of the columns of $\mathbf{V}=[\mathbf{v}_1,...,\mathbf{v}_N]$, or, equivalently, where the vector $\widehat{\bbx}=\mathbf{V}^{-1}\bbx$ is sparse \cite{SamplingOrtegaICASSP14}. In this context, vectors $\mathbf{v}_i$ are interpreted as the graph frequency basis, $\widehat{x}_i$ as the corresponding signal frequency coefficients, and $\bbx$ as a bandlimited graph signal. We assume that the set of active frequencies are known and, without loss of generality, that those are the first $K$ ones associated with the eigenvalues of largest magnitude \cite{SamplingOrtegaICASSP14,RabICASSP14_SpectCharSigsSmallWord}. Under this assumption, if we denote by $\widehat{\bbx}_K:=[\widehat{x}_1,\ldots,\widehat{x}_K]^T$ a $K \times 1$ vector collecting the coefficients corresponding to those frequencies, it holds that $\bbx$ is a $K$-bandlimited signal if
\begin{align}\label{E:signal_sparse_freq}
\widehat{\bbx}=[\widehat{\bbx}^T_K,0,\ldots,0]^T, \qquad \bbx=\mathbf{V}\widehat{\bbx}:=\mathbf{V}_K\widehat{\bbx}_K,
\end{align}
where we have defined the tall matrix $\mathbf{V}_K:=[\mathbf{v}_1,...,\mathbf{v}_K]$ containing the first $K$ eigenvectors of the shift operator $\bbS$.

\subsection{Graph filters}\label{Ss:graph_filters}

Graph filters $\mathbf{H}:\;\mathbb{R}^N \to \mathbb{R}^N$ are linear graph-signal operators of the form $\mathbf{H}:=\sum_{l=0}^{L-1}h_l \bbS^l$; i.e., polynomials (of degree $L-1$) of the graph-shift operator \cite{SandryMouraSPG_TSP13}. A particularity of graph filters is that they can be implemented locally, e.g., with $L-1$ exchanges of information among neighbors. This is true because the application of $\bbS$ on a signal $\bbx$ can be computed through local interactions.

The graph filter $\mathbf{H}$ can also be written as $\mathbf{H}=\mathbf{V}\big(\sum_{l=0}^{L-1}h_l \boldsymbol{\Lambda}^l\big) \mathbf{V}^{-1}$. The diagonal matrix $\widehat{\mathbf{H}}:=\sum_{l=0}^{L-1}h_l \boldsymbol{\Lambda}^l$ can then be viewed as the frequency response of $\mathbf{H}$ and it can be alternatively written as $\widehat{\mathbf{H}}=\diag{(\widehat{\mathbf{h}})}$, where vector $\widehat{\mathbf{h}}$ is a vector that contains the frequency response of the filter. Let $\lambda_i$ denote the $i$-th eigenvalue of $\bbS$ and define the $N \times L$ Vandermonde matrix
\begin{equation}\label{E:def_Psi_Vander}
\boldsymbol{\Psi}:= \left( \begin{array}{cccc}
1 & \lambda_1 & \ldots & \lambda_1^{L-1} \\
\vdots & \vdots &  &\vdots\\
1 & \lambda_N &  \ldots  & \lambda_N^{L-1} \end{array} \right).
\end{equation}
Upon defining the vector containing the coefficients of the filter as $\mathbf{h}:=[h_0,\ldots,h_{L-1}]^T$, it holds that $\widehat{\mathbf{h}}=\boldsymbol{\Psi}\mathbf{h}$ and therefore
\begin{equation}\label{E:Filter_from_time_to_freq}
\mathbf{H}=\!{\textstyle \sum_{l=0}^{L-1}h_l \bbS^l}=\!\mathbf{V}\diag\big(\boldsymbol{\Psi}\mathbf{h}\big) \!\mathbf{V}^{-1}\!=\!\mathbf{V}\diag(\widehat{\mathbf{h}})\mathbf{V}^{-1}\!.
\end{equation}
This implies that if $\bby$ is defined as $\bby=\mathbf{H}\bbx$, its frequency representation $\widehat{\bby}$ satisfies
\begin{equation}\label{E:Filter_input_output_freq}
\widehat{\bby}=\diag\big(\boldsymbol{\Psi}\mathbf{h}\big)\widehat{\bbx}.
\end{equation}
Within this context, a \emph{low-pass} graph filter of bandwidth $K$ is one where the frequency response $\widehat{\bbh} := \boldsymbol{\Psi}\mathbf{h}$ is given by
\begin{equation}\label{eqn_low_pass_filter}
\widehat{\bbh} = [ \widehat{\bbh}^T_K, 0, \ldots, 0]^T,
\end{equation}
where $\widehat{\bbh}_K$ contains the frequency response for the first $K$ frequencies. Notice that when the low-pass filter in \eqref{eqn_low_pass_filter} is applied to an arbitrary signal $\bbx$, the output signal is $K$-bandlimited as described in \eqref{E:signal_sparse_freq}. An alternative expression to define a graph filter is \cite{SuccNullingEigenv}
\begin{equation}\label{E:graph_filter_as_product}
\mathbf{H}= a_0 \prod_{l=1}^{L-1}(\bbS-a_l\mathbf{I}),
\end{equation}
which also gives rise to a polynomial on $\bbS$ of degree $L-1$. A specific advantage of the representation in \eqref{E:graph_filter_as_product} is that it provides a straightforward way to design low-pass filters via successive annihilation of graph frequencies. In particular, if we fix $a_l = \lambda_k$ for some eigenvalue $\lambda_k$ of $\bbS$ then the filter $\bbH$ will eliminate the frequency basis $\mathbf{v}_k$, i.e., the eigenvector associated with $\lambda_k$.
For future reference, we denote by $D$ the number of distinct eigenvalues in $\{\lambda_k\}_{k=K+1}^N$.
%


\begin{remark}[Discrete-time signals]\normalfont\label{R:time_domain}
To establish connections with classical time-varying signals, we define the directed cycle graph $\mathcal{G}_{dc}$, with node set $\ccalN = \{1, 2, \ldots, N\}$ and edge set $\mathcal{E}_{dc}\!=\!\{(i,\mymod(i)+1)\}_{i=1}^N$, where $\mymod(i)$ denotes the remainder obtained after dividing $i$ by $N$. Its adjacency and Laplacian matrices are denoted, respectively, as $\bbA_{dc}$ and $\mathbf{L}_{dc}\!:= \!\bbI\! -\! \bbA_{dc}$.
Discrete-time periodic signals can be thought as graph signals on the directed cycle $\mathcal{G}_{dc}$. Setting the shift operator either to $\bbS=\bbA_{dc}$ or $\bbS=\mathbf{L}_{dc}$ gives rise to the Fourier basis $\mathbf{F}$. More formally, the right eigenvectors of $\bbS$ satisfy $\mathbf{V}=\mathbf{F}$, with $F_{ij}:=\exp({+\mathfrak{j}2\pi(i-1)(j-1)/N})/\sqrt{N}$ where $\mathfrak{j} := \sqrt{-1}$. Selecting $\bbS=\bbA_{dc}$ has the additional advantage of satisfying $\Lambda_{ii}=\exp({-\mathfrak{j}2\pi(i-1)/N})$, i.e., the eigenvalues of the shift operator correspond to the classical discrete frequencies. Interpretations for the eigenvalues of $\mathbf{L}_{dc}$ also exist \cite{EmergingFieldGSP}.
The frequency representation $\widehat{\bbx}$ of a graph signal $\bbx$ is given by $\widehat{\bbx} = \bbV^{-1} \bbx$ whereas the frequency response of a filter with coefficients $\bbh$ is given by $\widehat{\bbh} = \boldsymbol{\Psi} \bbh$. For general graphs, matrices $\bbV^{-1}$ and $\boldsymbol{\Psi}$ need not be related. However, for the case of $\mathcal{G}_{dc}$, if $\bbS_{dc}=\bbA_{dc}$, then $\boldsymbol{\Psi} = \sqrt{N}\mathbf{F}^H$ and $\bbV^{-1} = \bbF^H$. Thus, the Fourier transforms for signals and filter coefficients are equivalent up to a constant for time-varying signals but this is not true for general graph signals.
\end{remark}

\subsection{Signal reconstruction using graph filters}\label{S:signal_reconstruction_graph_filters}

Our objective is to reconstruct a specific $K$-bandlimited signal $\bby$ by applying a graph filter $\bbH$ to a signal $\bbx$, where $\bbx$ is the result of a seeding procedure. More specifically, the reconstruction scheme proceeds in two phases:
\vspace{0.05in}
\begin{itemize}
\item{\emph{Seeding phase}.} The input to this phase is a set of $\tau$ graph signals $\{\bbs^{(t)}\}_{t=0}^{\tau-1}$, denominated seeding signals. These signals percolate through the graph following the dynamics
\begin{equation}\label{eqn_signal_y_time_t}
\bbx^{(t)} = \bbS \bbx^{(t-1)} + \bbs^{(t)}, \qquad\quad \bbx^{(-1)} = {\bf 0}.
\end{equation}
The output of this phase is set as $\bbx:=\bbx^{(\tau-1)}$.
\item{\emph{Filtering phase}.} The graph signal $\bbx$ is used as input to a low-pass graph filter $\mathbf{H}$, generating the output $\bbz:=\mathbf{H}\bbx$.
\end{itemize}
\vspace{0.05in}
\noindent The purpose of the seeding phase, which has duration $\tau$, is to inject into the graph the information needed to interpolate the signal $\bby$. The filtering phase further propagates the information available from the seeding phase while annihilating the frequencies with indices $k>K$ that are present in $\bbx$ but not in $\bby$. This phase has duration $L-1$, which is the order of the filter $\bbH$.

The goal of this paper is to design $\{\bbs^{(t)}\}_{t=0}^{\tau-1}$ and $\mathbf{H}$ such that $\bbz=\bby$. In Sections \ref{S:Extrapolating_GS_subset_nodes}, \ref{S:Extrapolating_GS_succesive}, and \ref{S:shift_space_extrapolation} we present this design for three different seeding schemes, where we impose additional restrictions on the structure and the number of seeding signals.

\begin{remark}\normalfont
{In classical discrete-time signal processing, recovery of bandlimited signals is a two-step process. Firstly, a sparse regular signal whose non-zero values coincide with those of the signal to recover is generated. Secondly, the (zero) values not specified in the sparse signal are extrapolated from the non-zero ones using a low-pass filter. Our approach in this paper is to preserve this two-step methodology and use it to recover bandlimited graph signals. This provides a way to regenerate a desired signal in a graph -- either estimated from samples or otherwise --€" by acting on a subset of (seeding) nodes. As it will be shown in the ensuing sections, for signals defined on a general graph, the non-zero values of the sparse signal in the first step will not coincide with those of the signal to recover.
This deviates from the classical concept of interpolation, which assumes that the non-zero values are the same than those of the signal to reconstruct.}

{The practical advantage of studying recovery schemes that use graph filters is twofold. First, they can be implemented distributedly, using only local exchanges among neighbors. Second, since \emph{graph filters} can be used to \emph{model} diffusion processes (e.g. the spread of an opinion in a social network), our results can be used to reconstruct signals in network applications that implement linear diffusion dynamics.}
\end{remark}

\section{Multiple node - single time seeding}\label{S:Extrapolating_GS_subset_nodes}

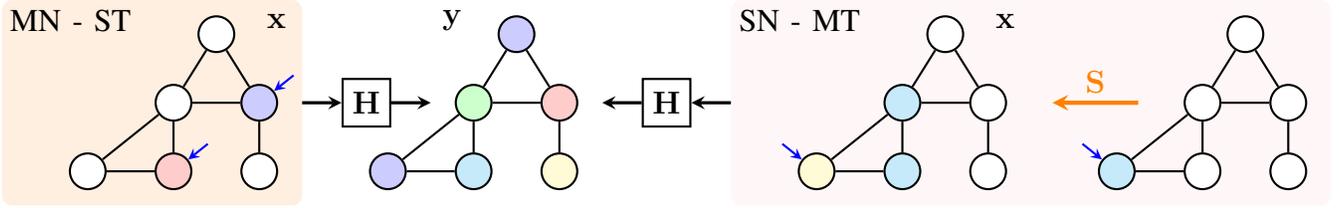
\begin{figure*}
\centering
\def \thisplotscale {0.95}
\def \unit {\thisplotscale cm}
\tikzstyle {blue vertex here} = [blue vertex, 
                                 minimum width = 0.5*\unit, 
                                 minimum height = 0.5*\unit, 
                                 anchor=center,
                                 font=\scriptsize]
\tikzstyle {vertex here} = [vertex, 
                                 minimum width = 0.5*\unit, 
                                 minimum height = 0.5*\unit, 
                                 anchor=center,
                                 font=\scriptsize]
\tikzstyle {red vertex here} = [red vertex, 
                                 minimum width = 0.5*\unit, 
                                 minimum height = 0.5*\unit, 
                                 anchor=center,
                                 font=\scriptsize]
\tikzstyle {green vertex here} = [green vertex, 
                                 minimum width = 0.5*\unit, 
                                 minimum height = 0.5*\unit, 
                                 anchor=center,
                                 font=\scriptsize]
\tikzstyle {yellow vertex here} = [yellow vertex, 
                                 minimum width = 0.5*\unit, 
                                 minimum height = 0.5*\unit, 
                                 anchor=center,
                                 font=\scriptsize]
\tikzstyle {cyan vertex here} = [cyan vertex, 
                                 minimum width = 0.5*\unit, 
                                 minimum height = 0.5*\unit, 
                                 anchor=center,
                                 font=\scriptsize]
\tikzstyle {black vertex here} = [black vertex, 
                                 minimum width = 0.5*\unit, 
                                 minimum height = 0.5*\unit, 
                                 anchor=center,
                                 font=\scriptsize]
\tikzstyle {white vertex here} = [white vertex, 
                                 minimum width = 0.5*\unit, 
                                 minimum height = 0.5*\unit, 
                                 anchor=center,
                                 font=\scriptsize]
\tikzstyle{int}=[draw, fill=white!20, minimum size=2em]
{\small \begin{tikzpicture}[thick, x = 1.2*\unit, y = 0.96*\unit]

	\small 
	
	\node at (-0.5,0) (center1) {};
	\path (center1) ++ (0,0) node (1) [green vertex here] {};
	\path (center1) ++ (1,0) node (2) [red vertex here] {};
	\path (center1) ++ (0.5,1) node (3) [blue vertex here] {};
	\path (center1) ++ (-1, -1) node (4) [blue vertex here] {};
	\path (center1) ++ (0,-1) node (5) [cyan vertex here] {};
	\path (center1) ++ (1,-1) node (6) [yellow vertex here] {};

	\path[thick] (1) edge [] node {{}} (2);
	\path[thick] (1) edge [] node {{}} (3);		                            
	\path[thick] (1) edge [] node {{}} (4);		                            
	\path[thick] (1) edge [] node {{}} (5);		                            
	\path[thick] (2) edge [] node {{}} (3);		                            
	\path[thick] (2) edge [] node {{}} (6);		                            
	\path[thick] (4) edge [] node {{}} (5);

		
	\fill[orange!30, fill opacity = 0.4, rounded corners] (-4-2, 0-1.5) rectangle (-4+1.5, 0+1.5);
	
	\node at (-4,0) (center2) {};
	\path (center2) ++ (0,0) node (1p) [white vertex here] {};
	\path (center2) ++ (1,0) node (2p) [blue vertex here] {};
	\path (center2) ++ (0.5,1) node (3p) [white vertex here] {};
	\path (center2) ++ (-1, -1) node (4p) [white vertex here] {};
	\path (center2) ++ (0,-1) node (5p) [red vertex here] {};
	\path (center2) ++ (1,-1) node (6p) [white vertex here] {};

	\path[thick] (1p) edge [] node {{}} (2p);
	\path[thick] (1p) edge [] node {{}} (3p);		                            
	\path[thick] (1p) edge [] node {{}} (4p);		                            
	\path[thick] (1p) edge [] node {{}} (5p);		                            
	\path[thick] (2p) edge [] node {{}} (3p);		                            
	\path[thick] (2p) edge [] node {{}} (6p);		                            
	\path[thick] (4p) edge [] node {{}} (5p);	
	
	\node at (-4+2.25,0) [int] (filter1) {{\large $\bbH$}};
	\path[very thick, -stealth] (-4+1.5, 0) edge [] node {{}} (filter1);
	\path[very thick, -stealth] (filter1) edge [] node {{}} (-4+2.25+0.75, 0);

	
	\path[thick, -stealth] (2p)+(0.4,0.4) edge [blue] node {{}} (2p);
	\path[thick, -stealth] (5p)+(0.4,0.4) edge [blue] node {{}} (5p);

	
	\fill[pink!30, fill opacity = 0.4, rounded corners] (4-1.5, 0-1.5) rectangle (8+1.5, 0+1.5);	
		
	\node at (8,0) (center3) {};
	\path (center3) ++ (0,0) node (1pp) [white vertex here] {};
	\path (center3) ++ (1,0) node (2pp) [white vertex here] {};
	\path (center3) ++ (0.5,1) node (3pp) [white vertex here] {};
	\path (center3) ++ (-1, -1) node (4pp) [cyan vertex here] {};
	\path (center3) ++ (0,-1) node (5pp) [white vertex here] {};
	\path (center3) ++ (1,-1) node (6pp) [white vertex here] {};

	\path[thick] (1pp) edge [] node {{}} (2pp);
	\path[thick] (1pp) edge [] node {{}} (3pp);		                            
	\path[thick] (1pp) edge [] node {{}} (4pp);		                            
	\path[thick] (1pp) edge [] node {{}} (5pp);		                            
	\path[thick] (2pp) edge [] node {{}} (3pp);		                            
	\path[thick] (2pp) edge [] node {{}} (6pp);		                            
	\path[thick] (4pp) edge [] node {{}} (5pp);	
	
	\path[thick, -stealth] (4pp)+(-0.4,+0.4) edge [blue] node {{}} (4pp);
	
	\path[ultra thick, -stealth] (8-0.75, 0) edge [orange, above] node {\large {$\bbS$}} (8-1.75, 0);

	\node at (4.5,0) (center4) {};
	\path (center4) ++ (0,0) node (1ppp) [cyan vertex here] {};
	\path (center4) ++ (1,0) node (2ppp) [white vertex here] {};
	\path (center4) ++ (0.5,1) node (3ppp) [white vertex here] {};
	\path (center4) ++ (-1, -1) node (4ppp) [yellow vertex here] {};
	\path (center4) ++ (0,-1) node (5ppp) [cyan vertex here] {};
	\path (center4) ++ (1,-1) node (6ppp) [white vertex here] {};

	\path[thick] (1ppp) edge [] node {{}} (2ppp);
	\path[thick] (1ppp) edge [] node {{}} (3ppp);		                            
	\path[thick] (1ppp) edge [] node {{}} (4ppp);		                            
	\path[thick] (1ppp) edge [] node {{}} (5ppp);		                            
	\path[thick] (2ppp) edge [] node {{}} (3ppp);		                            
	\path[thick] (2ppp) edge [] node {{}} (6ppp);		                            
	\path[thick] (4ppp) edge [] node {{}} (5ppp);

	\node at (4-2.25,0) [int] (filter2) {{\large $\bbH$}};
	\path[very thick, -stealth] (4-1.5, 0) edge [] node {{}} (filter2);
	\path[very thick, -stealth] (filter2) edge [] node {{}} (4-2.25-0.75, 0);
	
	
	\path[thick, -stealth] (4ppp)+(-0.4,+0.4) edge [blue] node {{}} (4ppp);

	\large
	  \node at (-0.75, 1.2)  {$\bby$};
	 \path (center2) ++ (1.2,1.2) node () [] {$\bbx$};
	 \path (center2) ++ (-1.2,1.2) node () [] {MN - ST};
	 \path (center4) ++ (1.2,1.2) node () [] {$\bbx$};
	 \path (center4) ++ (-1.2,1.2) node () [] {SN - MT};

\end{tikzpicture}}
\caption{{Two different schemes to reconstruct signal $\bby$. On the left, MN-ST seeding injects $P=2$ values simultaneously (blue arrows), after which the low-pass filter $\bbH$ is applied. On the right, SN-MT seeding first injects a single value that percolates to the two neighboring nodes. After a second value injection at the same node, filter $\bbH$ completes the reconstruction.}}
\label{F:example_seedings}
\end{figure*}

In multiple node - single time (MN-ST) seeding we consider the particular case where there is only $\tau = 1$ seeding signal $\bbs$ so that $\bbx = \bbs$ [cf. \eqref{eqn_signal_y_time_t}]. Denoting by $P$ the amount of nonzero values in $\bbs$, we interpret MN-ST seeding as having $P$ seeding nodes that inject a \emph{single} value, while the remaining $N-P$ nodes keep silent{; see left and center panels in Figure~\ref{F:example_seedings}}. Define the signal injected by node $i$ as $s_i$ and assume, without loss of generality, that the seeding nodes are the $P$ first ones. We therefore define the $ P\times 1$ and $N \times 1$ seeding vectors as
\begin{align}
\bbs_P&=[s_1,\ldots,s_P]^T, \\
\bbs&=[s_1,\ldots,s_P, 0,\ldots, 0]^T.\label{E:def_x_bar}
\end{align}
Then, given a bandlimited signal $\bby=\mathbf{V}_K \widehat{\bby}_K$ [cf.~\eqref{E:signal_sparse_freq}], our goal is to design $\mathbf{H}$ and $\bbs$ such that
\begin{equation}\label{E:goal_time_setup1}
\bby=\mathbf{H}\bbs,
\end{equation}
where $\mathbf{H}$ has the \emph{particular structure of a graph filter} (cf.~Section~\ref{Ss:graph_filters}).
Exploiting the fact that $\bby$ is bandlimited, it is reasonable to write \eqref{E:goal_time_setup1} in the frequency domain. To do this, both sides of \eqref{E:goal_time_setup1} are left multiplied by $\mathbf{V}^{-1}$, which yields
\begin{equation}\label{E:goal_time_setup2}
\widehat{\bby}\!=\!\mathbf{V}^{-1}\mathbf{H}\bbs=\mathbf{V}^{-1}\mathbf{V}\diag(\boldsymbol{\Psi}\mathbf{h})\mathbf{V}^{-1}\bbs=\diag(\boldsymbol{\Psi}\mathbf{h})\widehat{\bbs},
\end{equation}
where we used \eqref{E:Filter_from_time_to_freq} for the second equality. Utilizing the fact that the seeding signal $\bbs$ is sparse [cf.~\eqref{E:def_x_bar}] we may write its frequency representation as
\begin{equation}\label{E:seding_sig_freq_dom_setup1}
\widehat{\bbs}=\mathbf{V}^{-1}\bbs=\mathbf{V}^{-1}\bbE_P\bbs_P,
\end{equation}
where, {we recall}, $\bbE_P:=[\bbe_1,...,\bbe_P]$ is a tall matrix collecting the $P$ first canonical basis vectors of size $N \times 1$.
By substituting \eqref{E:seding_sig_freq_dom_setup1} into \eqref{E:goal_time_setup2}, our goal of designing $\mathbf{H}$ and $\bbs$ such that $\bby=\mathbf{H}\bbs$ can be reformulated as designing $\mathbf{h}$ and $\bbs_P$ such that
\begin{equation}\label{E:goal_freq_setup1}
\widehat{\bby}=\diag\big(\boldsymbol{\Psi}\mathbf{h}\big)\mathbf{V}^{-1}\bbE_P\bbs_P,
\end{equation}
which is a bilinear system of $N$ equations and $L+P$ variables.
Leveraging the sparsity of $\widehat{\bby}$ [cf.~\eqref{E:signal_sparse_freq}], the system of $N$ equations in \eqref{E:goal_freq_setup1} can be split into two
\begin{equation}\label{E:goal_freq_setup1_split_2_nonzeros}
\widehat{\bby}_K = \bbE_K^T\, \diag(\boldsymbol{\Psi}\mathbf{h})\, \mathbf{V}^{-1}\bbE_P\bbs_P,
\end{equation}
\begin{equation}\label{E:goal_freq_setup1_split_2_zeros}
\mathbf{0}_{N - K} = \bar{\bbE}_K^T \, \diag(\boldsymbol{\Psi}\mathbf{h}) \, \mathbf{V}^{-1}\bbE_P \bbs_P,
\end{equation}
where, {we recall}, $\bar{\bbE}_K:=[\bbe_{K+1},...,\bbe_N]$ collects the last $N-K$ canonical basis vectors and $\mathbf{0}_{N - K}$ denotes the $(N-K) \times 1$ vector of all zeros.
Note that the conditions in \eqref{E:goal_freq_setup1_split_2_zeros} are the same for any $K$-bandlimited signal. On the other hand, the conditions in \eqref{E:goal_freq_setup1_split_2_nonzeros} depend on the specific signal to be interpolated. A natural approach is to use the filter coefficients $\mathbf{h}$ -- which are related to the global behavior of the graph -- to guarantee that \eqref{E:goal_freq_setup1_split_2_zeros} holds, while using the seeding signal $\bbs_P$ to satisfy \eqref{E:goal_freq_setup1_split_2_nonzeros} and, hence, to guarantee that the output of the interpolation is $\bby$. In this way, the filter coefficients $\bbh$ to be designed do not depend on the particular signal to reconstruct.

The conditions under which the mentioned approach is guaranteed to find a feasible solution are given in the form of two propositions. Ensuing discussions describe the actual procedure to interpolate the signal.
%
\begin{myproposition}\label{P:Recov_filter_setup1}
If $L>D$ (cf.~Section~\ref{Ss:graph_filters}), there exist infinitely many nonzero $L \times 1$ vectors $\mathbf{h}^*$ such that, after setting $\mathbf{h}=\mathbf{h}^*$, \eqref{E:goal_freq_setup1_split_2_zeros} is satisfied for any $\bbV^{-1}\!$ and $\bbs_P$.
\end{myproposition}
\begin{myproof}
Since \eqref{E:goal_freq_setup1_split_2_zeros} has to hold for any seeding signal $\bbs_P$, we need $\bar{\bbE}_K^T\boldsymbol{\Psi}\mathbf{h}=\mathbf{0}$. This requires $\mathbf{h}$ to belong to the kernel of the $(N-K)\times L$ matrix $\bar{\bbE}_K^T\boldsymbol{\Psi}$. Since $\bar{\bbE}_K^T\boldsymbol{\Psi}$ is a Vandermonde matrix, its number of linearly independent rows is equal to the number of distinct eigenvalues in $\{\lambda_k\}_{k=K+1}^N$, which is $D$. Thus, the existence of a solution $\mathbf{h}^*\neq \mathbf{0}$ requires $L>D$.
\end{myproof}

For $L>D$, $\bar{\bbE}_K^T\boldsymbol{\Psi}$ is rank deficient and the dimension of its kernel space is $L-D$. Hence, setting $\mathbf{h}$ to any nonzero element of the kernel space will satisfy  \eqref{E:goal_freq_setup1_split_2_zeros}. In what follows, we will assume that $L=D+1$ and set the coefficients $\mathbf{h}^*$ that solve \eqref{E:goal_freq_setup1_split_2_zeros} as the unit vector spanning the unidimensional kernel space of $\bar{\bbE}_K^T\boldsymbol{\Psi}$. For the case where all the eigenvalues $\{\lambda_k\}_{k=K+1}^N$ are distinct, this implies that $L=N-K+1$.

Once the coefficients of the filter are designed, the next step is to find the optimum seeding signal. With  $\widehat{\bbh}_K^* := \bbE_K^T \, \boldsymbol{\Psi}\mathbf{h}^*$ denoting the frequency response of the low-pass filter in the active frequencies, substituting $\mathbf{h}=\mathbf{h}^*$ into \eqref{E:goal_freq_setup1_split_2_nonzeros} yields
\begin{equation}\label{E:goal_freq_setup1_split_2_nonzeros_after_filter}
\widehat{\bby}_K = \diag(\widehat{\bbh}_K^*)\, \bbE_K^T \mathbf{V}^{-1}\bbE_P\bbs_P.
\end{equation}
For which the following result holds.
%
\begin{myproposition}\label{P:Recov_seeds_setup1}
The system of $K$ equations in \eqref{E:goal_freq_setup1_split_2_nonzeros_after_filter} is guaranteed to have a solution with respect to $\bbs_P$ if the following two conditions hold:\\
 i) $\lambda_{k_1}\neq\lambda_{k_2}$ for all $(\lambda_{k_1},\lambda_{k_2})$ such that $k_1 \leq K$ and $k_2>K$, \\
 ii) $\mathrm{rank}(\bbE_K^T \mathbf{V}^{-1}\bbE_P) \geq K$.
\end{myproposition}
\begin{myproof}
Condition~\emph{i)} is required to guarantee that all the elements of $\widehat{\bbh}_K^*$ are nonzero.
We prove this by contradiction.
Recall that the following facts holds true (cf. Proposition~\ref{P:Recov_filter_setup1}): a) $\bar{\bbE}_K^T\boldsymbol{\Psi}\mathbf{h}^*=\mathbf{0}_{N-K}$; b) $\mathbf{h}^*\neq\mathbf{0}_L$ and c) rank of $\bar{\bbE}_K^T\boldsymbol{\Psi}$ is $L-1$. Assume, without loss of generality, that the element of $\widehat{\mathbf{h}}_K^*$ that is zero is the $K$-th one. Then, we can use a) to write $\bar{\bbE}_{K-1}^T\boldsymbol{\Psi}\mathbf{h}^*=\mathbf{0}_{N-K+1}$. Condition \emph{i)} and fact c) guarantee that $\bar{\bbE}_{K-1}^T\boldsymbol{\Psi}$ has rank $L$; then, satisfying $\bar{\bbE}_{K-1}^T\boldsymbol{\Psi}\mathbf{h}^*=\mathbf{0}_{N-K+1}$ requires $\mathbf{h}^*=\mathbf{0}_L$, which contradicts b). Hence, all the elements of $\widehat{\bbh}_K^*$ are nonzero. This guarantees that $\diag(\widehat{\bbh}_K^*)$ is invertible, so that \eqref{E:goal_freq_setup1_split_2_nonzeros_after_filter} can be written as
%
%
%
%

%
\begin{equation}\label{E:goal_freq_setup1_split_2_nonzeros_after_filter_inv}
\diag(\widehat{\bbh}_K^*)^{-1} \widehat{\bby}_K = (\bbE_K^T \mathbf{V}^{-1}\bbE_P) \bbs_P,
\end{equation}
where $\bbE_K^T \mathbf{V}^{-1}\bbE_P$ is a $K \times P$ submatrix of $\mathbf{V}^{-1}$. To guarantee that the system of equations in \eqref{E:goal_freq_setup1_split_2_nonzeros_after_filter_inv} has at least one solution, we need condition~\emph{ii)} to hold.
\end{myproof}

Different from the time domain, where all the eigenvalues of $\bbS$ (frequencies) are distinct, in the more general graph domain there can be graph topologies that give rise to $\bbS$ with repeated eigenvalues. Condition~\emph{i)} is required because a graph filter $\mathbf{H}$ always produces the same frequency response if the corresponding eigenvalues are the same. Therefore, it is not possible for $\mathbf{H}$ to eliminate one of the frequencies without eliminating the other. An alternative to bypass this problem is discussed in Section~\ref{Ss:low_pass_extrap_augmented}. Condition~\emph{ii)} requires the rank of the $K\times P$ matrix $\bbE_K^T \mathbf{V}^{-1}\bbE_P$ being at least $K$. At the very least, this requires $P$, the number of seeding nodes, to be equal to $K$, the number of frequencies present in $\bby$. For the particular case of $P=K$, if the conditions in the proposition are satisfied, the $\bbs_P$ that recovers $\bby$ is
\begin{equation}\label{E:goal_freq_setup1_split_2_nonzeros_after_filter_inv00}
\bbs_P = (\bbE_K^T \mathbf{V}^{-1}\bbE_P)^{-1} \diag(\widehat{\bbh}_K^*)^{-1} \widehat{\bby}_K.
\end{equation}
However, if condition ii) is not satisfied, there may be cases where setting $P=K$ fails. To see why this is true, notice that $[\mathbf{V}^{-1}]_{k,p}$ can be viewed as how strongly node $p$ expresses frequency $k$. Suppose for example that there exists a $k$ such that $[\mathbf{V}^{-1}]_{k,p}=0$ for all nodes $p=1,\ldots,P$, then it is not possible to reconstruct a signal $\bby$ with $\widehat{\bby}_k\neq 0$ using that set of nodes. This problem is also present when sampling graph signals by observing the value of the signal in a subset of nodes \cite{SamplingKovacevic_without_Moura_15}.

Proposition~\ref{P:Recov_seeds_setup1} states conditions for \emph{perfect} reconstruction in a noiseless setting. In noisy scenarios, the specific set of nodes selected to inject the seeding signal has an impact on the reconstruction error. This is analyzed in Section~\ref{sec_imperfect_reconstruction}.

\subsection{Filter degree reduction in MN-ST seeding}\label{Ss:low_pass_extrap_augmented}

The MN-ST reconstruction scheme requires a low-pass filter $\bbH$ of degree $D$, which grows with the size of the graph. Since the degree of $\mathbf{H}$ corresponds to the number of local interactions needed to implement the filter, the communication overhead can be a problem for large graphs. In this context, we look for solutions that reduce the degree of $\bbH$ by increasing the number of seeding nodes $P$. This can be done by splitting the system of equations in \eqref{E:goal_freq_setup1} as [cf.~\eqref{E:goal_freq_setup1_split_2_nonzeros}-\eqref{E:goal_freq_setup1_split_2_zeros}]
\begin{equation}\label{E:goal_freq_setup1_split_2_nonzeros_2}
[\widehat{\bby}^T_K, \mathbf{0}^T_{P-K}]^T = \bbE_P^T\, \diag(\boldsymbol{\Psi}\mathbf{h})\, \mathbf{V}^{-1}\bbE_P\bbs_P,
\end{equation}
\begin{equation}\label{E:goal_freq_setup1_split_2_zeros_2}
\mathbf{0}_{N - P} = \bar{\bbE}_P^T \, \diag(\boldsymbol{\Psi}\mathbf{h}) \, \mathbf{V}^{-1}\bbE_P \bbs_P\!.
\end{equation}
The filter coefficients must be obtained now to annihilate the $N-P$ frequencies in \eqref{E:goal_freq_setup1_split_2_zeros_2} and the $P$ seeding nodes must inject a signal whose spectrum, after being filtered by $\mathbf{H}$, matches that of the desired signal \eqref{E:goal_freq_setup1_split_2_nonzeros_2}. Notice that \eqref{E:goal_freq_setup1_split_2_nonzeros_2}-\eqref{E:goal_freq_setup1_split_2_zeros_2} can also be used when \eqref{E:goal_freq_setup1_split_2_nonzeros}-\eqref{E:goal_freq_setup1_split_2_zeros} fail due to a violation of condition~\emph{i)} in Proposition~\ref{P:Recov_seeds_setup1}. More specifically, for every frequency index $k_2 > K$ with the same eigenvalue as a frequency index $k_1 \leq K$ we can induce a zero frequency coefficient in the reconstructed signal via the seeding values [cf.~\eqref{E:goal_freq_setup1_split_2_nonzeros_2}] instead of through the low-pass filter [cf.~\eqref{E:goal_freq_setup1_split_2_zeros_2}] and, hence, drop condition~\emph{i)} as a requirement for recovery. Further notice that for \eqref{E:goal_freq_setup1_split_2_zeros_2} to hold for any $\bbs_P$, the degree of the filter needs to be at least equal to the number of distinct eigenvalues in $\{\lambda_k\}_{k=P+1}^N$ (cf.~Proposition~\ref{P:Recov_filter_setup1}). In the extreme case of $P=N$, the trivial solution $\mathbf{h}=[1,0,\ldots,0]^T$ (0-order filter) and $\bbs_P = \bby$ satisfies \eqref{E:goal_freq_setup1_split_2_nonzeros_2}-\eqref{E:goal_freq_setup1_split_2_zeros_2}.


\subsection{Relation to classical interpolation}\label{Ss:Compar_with_low_pass_sinc_interp}

In the classical time domain, sinc (low-pass) interpolation of a bandlimited signal leads to perfect reconstruction. If the sampling is performed at the minimum possible rate, the bandwidth of the low-pass filter has to be exactly the same than that of $\bby$. By contrast, if the signal is oversampled, the bandwidth can be larger. Equivalently, if more samples than the minimum required number are available, then the low-pass filter does not have to cancel all the frequencies not present in $\bby$. The analysis in Section~\ref{Ss:low_pass_extrap_augmented} reveals that this is also the case when signals are defined in more general graph domains.

The main differences between the MN-ST reconstruction scheme and classical time interpolation come from the fact that the basis $\bbV$ of a general graph shift $\bbS$ is not as structured as the Fourier basis $\bbF$. A difference of particular relevance is that, for general graphs, the seeding values $\bbs_P$ do not coincide with the values of the desired signal $\bby$. This contrasts with the classical interpolation of uniformly sampled time-varying signals, where $\bbs_P$ is a subset of the signal $\bby$.
In fact, it can be rigorously shown that requiring such a condition for general graphs would lead to an infeasible interpolation.
To be concrete, suppose that $P=K$, so that $\bbs=[\bbs_K^T,\mathbf{0}^T]^T$ [cf. \eqref{E:def_x_bar}], and that $\bbs_K$ is equal to the first entries of $\bby$. We can then leverage the fact that $\bbs$ and $\widehat{\bby}$ are sparse to write
\begin{equation}\label{E:intep_graph_as_time_cond1}
\bbs_K=\bbE_K^T \bby=\bbE_K^T \mathbf{V}\widehat{\bby}=\bbE_K^T \mathbf{V}\bbE_K\widehat{\bby}_K.
\end{equation}
Secondly, we write the goal of $\bby=\mathbf{H}\bbs$ into the frequency domain as $\widehat{\bby}=\diag{(\widehat{\mathbf{h}})}\mathbf{V}^{-1}\bbs$ and use again the sparsity of $\bbs$ and $\widehat{\bby}$ to write
\begin{align}\label{E:intep_graph_as_time_cond2}
\widehat{\bby}_{K}
&\!=\!\bbE_{K}^T\diag{(\widehat{\mathbf{h}})}\mathbf{V}^{-1}\bbE_{K}\bbs_K
\!=\!\bbE_{K}^T\diag{(\widehat{\mathbf{h}})}\bbE_{K}\bbE_{K}^T\mathbf{V}^{-\!1}\bbE_{K}\bbs_K\nonumber\\
&=\diag{(\widehat{\mathbf{h}}_K)}\bbE_{K}^T\mathbf{V}^{-\!1}\bbE_{K}\bbs_K,
\end{align}
where $\widehat{\mathbf{h}}_K:=\bbE^T_{K} \widehat{\mathbf{h}}$ contains the first $K$ components of $\widehat{\mathbf{h}}$. Substituting \eqref{E:intep_graph_as_time_cond1} into \eqref{E:intep_graph_as_time_cond2} yields
\begin{equation}\label{E:intep_graph_as_time_cond3}
\widehat{\bby}_{K}=\diag{(\widehat{\mathbf{h}}_K)}\bbE_{K}^T\mathbf{V}^{-\!1}\bbE_{K} \bbE_K^T \mathbf{V}\bbE_K \widehat{\bby}_K.
\end{equation}
Since \eqref{E:intep_graph_as_time_cond3} must hold for all $\widehat{\bby}_K$, it can only be satisfied if $\diag{(\widehat{\mathbf{h}}_K)}\bbE_{K}^T\mathbf{V}^{-\!1}\bbE_{K} \bbE_K^T \mathbf{V}\bbE_K\!\!=\!\!\mathbf{I}$. This requires matrix $(\bbE_{K}^T\mathbf{V}^{-\!1}\bbE_{K} \bbE_K^T \mathbf{V}\bbE_K)$ to be diagonal. While this is true when $K=N$, it is not true for a general $K$. However, in the time domain where $\bbV = \bbF$ (see Remark~\ref{R:time_domain}), for some cases the multiplication of submatrices of $\mathbf{F}$ is guaranteed to be diagonal. For example, if the $K$ seeding nodes are chosen uniformly (equally) spaced, then $(\bbE_{K}^T\mathbf{V}^{-\!1}\bbE_{K} \bbE_K^T \mathbf{V}\bbE_K) = K/N \,\bbI$. This implies not only that \eqref{E:intep_graph_as_time_cond3} is satisfied, but also that all the entries in $\widehat{\mathbf{h}}_K$ must be set to $N/K$. In other words, the optimal low-pass interpolator after uniform sampling in the time domain has the same response for all the active frequencies, as known from classical signal processing.

\section{Single node - multiple time seeding}\label{S:Extrapolating_GS_succesive}

In single node - multiple time (SN-MT) seeding, we consider the particular case where all the $\tau = P$ seeding values are injected at a \emph{single} node{; see right and center panels in Figure~\ref{F:example_seedings}}. To be more specific, assume without loss of generality that the first node is the one injecting the seeding values, so that the seeding signal $\bbs^{(t)}$ at time $t$ is of the form $\bbs^{(t)}=[s^{(t)},0,\ldots,0]^T$. Then, define $\bbs_P:=[s^{(P-1)},\ldots,s^{(0)}]^T$ to be a $P \times 1$ vector grouping the seeding values.
We present the relation between the seeding values $\bbs_P$ and the output of the seeding phase $\bbx$ in the following lemma.
%
\begin{mylemma}\label{lem_seeding_phase}
The frequency representation of the intermediate signal $\bbx$ in SN-MT seeding is given by
\begin{align}\label{E:interp_shift_ver3_2}
\widehat{\bbx} = \diag(\widehat{\bbe}_1)\boldsymbol{\Psi} \bbs_P,
\end{align}
where $\widehat{\bbe}_1:=\mathbf{V}^{-1}\bbe_1$ is the frequency representation of the first canonical basis vector.
\end{mylemma}
\begin{myproof}
Since $\bbx$ is obtained after $P$ injections of seeding values following the dynamics in \eqref{eqn_signal_y_time_t}, it holds that
\begin{equation}\label{eqn_signal_y_time_t_2}
\bbx = \bbx^{(P-1)}={\textstyle \sum_{l=0}^{P-1} \bbS^l \bbs^{(P-1-l)} = \sum_{l=0}^{P-1} \bbS^l s^{(P-1-l)} \bbe_1}.
\end{equation}
Equation \eqref{eqn_signal_y_time_t_2} relates the signal $\bbx$ to the successive inputs $s^{(\cdot)}$ of the seeding node and can be interpreted as the application of the graph filter
\begin{equation}\label{eqn_interpolation_as_filtering}
\bar{\bbH} := {\textstyle\sum_{l=0}^{P-1} s^{(P-1-l)} \bbS^l}
\end{equation}
of degree $P-1$ to the canonical basis vector $\bbe_1$. Building on this interpretation, we may use \eqref{E:Filter_input_output_freq} to write
\begin{align}\label{E:interp_shift_ver3}
\widehat{\bbx} =\diag(\boldsymbol{\Psi}\bbs_P)\widehat{\bbe}_1,
\end{align}
and, by exploiting the fact that for generic vectors $\mathbf{a}$ and $\mathbf{b}$ it holds that $\diag(\mathbf{a})\mathbf{b}=\diag(\mathbf{b})\mathbf{a}$, the lemma follows.
\end{myproof}

\noindent The proof of Lemma~\ref{lem_seeding_phase} exploits the reinterpretation of the seeding phase as the application of a graph filter $\bar{\bbH}$, whose coefficients are given by the seeding values, to the canonical basis vector $\bbe_1$. Equation \eqref{E:interp_shift_ver3_2} reveals how $\widehat{\bbx}$ depends on the structure of the graph $\boldsymbol{\Psi}$ and the seeding values $\bbs_P$, as well as on the particular node chosen to inject the values via $\widehat{\bbe}_1$, whose elements represent how strongly the node expresses each of the graph frequencies.

The next step is to analyze the output of the filtering phase in the frequency domain $\widehat{\bbz}$. To do this, recall that $\bbh^*$ denotes the coefficients of a low-pass filter (cf. Section~\ref{Ss:graph_filters}) that eliminates all frequencies with indices $k > K$.
Defining $\widehat{\mathbf{h}}^*:=\boldsymbol{\Psi}\mathbf{h}^*$, we may analyze the application of the low-pass filter in the frequency domain as
\begin{equation}\label{E:one_seed_low_pas_phase2_a}
\widehat{\bbz}_K=\bbE_K^T\diag(\widehat{\mathbf{h}}^*) \widehat{\bbx} =\bbE_K^T\diag(\widehat{\mathbf{h}}^*)\bbE_K\bbE_K^T\widehat{\bbx}.
\end{equation}
Further recalling that $\widehat{\mathbf{h}}_K^*:=\bbE_K^T\widehat{\mathbf{h}}^*$ and substituting \eqref{E:interp_shift_ver3_2} into \eqref{E:one_seed_low_pas_phase2_a}, we obtain [cf.~\eqref{E:goal_freq_setup1_split_2_nonzeros_after_filter}]
\begin{equation}\label{eqn_solution_one_seed_low_pass}
\widehat{\bby}_K = \diag(\widehat{\mathbf{h}}_K^*) \bbE_K^T\diag(\widehat{\bbe}_1)\boldsymbol{\Psi}\bbs_P.
\end{equation}
Expression \eqref{eqn_solution_one_seed_low_pass} relates the frequencies present in $\bby$ to the seeding values $\bbs_P$. Provided that $K\leq P$, the following proposition states the conditions under which \eqref{eqn_solution_one_seed_low_pass} can be solved with respect to $\bbs_P$.
%
\begin{myproposition}\label{P:Recov_single_seeds_setup3}
Let $U_1$ be the number of values in $\{[\widehat{\bbe}_1]_k\}_{k=1}^{K}$ that are zero and let $D_1$ be the number of repeated values in $\{\lambda_k\}_{k=1}^{K}$. Then, the system of $K$ equations in \eqref{eqn_solution_one_seed_low_pass} is guaranteed to have a solution with respect to $\bbs_P$ if the following two conditions hold:\\
\noindent i) $\lambda_{k_1}\neq\lambda_{k_2}$ for all $(\lambda_{k_1},\lambda_{k_2})$ such that $k_1 \leq K$ and $k_2>K$,\\
\noindent ii) $U_1=0$ and $D_1=0$.
\end{myproposition}
\begin{myproof}
If we rewrite \eqref{eqn_solution_one_seed_low_pass} as
\begin{equation}\label{eqn_proof_prop_seeds_setup3_010}
\widehat{\bby}_K = \Big(\diag(\widehat{\mathbf{h}}_K^*)\Big) \left(\bbE_K^T\diag(\widehat{\bbe}_1)\bbE_K \right) \left( \bbE_K^T\boldsymbol{\Psi} \right) \bbs_P,
\end{equation}
then it becomes clear that conditions~\emph{i)} and~\emph{ii)} ensure invertibility of the two square matrices, and full row rank of the rectangular matrix $\bbE_K^T\boldsymbol{\Psi}$. To be specific, condition~\emph{i)} is required to guarantee that all the entries of vector $\widehat{\mathbf{h}}_K^*$ are nonzero and, hence, matrix $\diag(\widehat{\mathbf{h}}_K^*)$ is invertible (cf. proof of Proposition~\ref{P:Recov_seeds_setup1}). Condition $U_1=0$ in \emph{ii)} ensures that $\bbE_K^T\diag(\widehat{\bbe}_1)\bbE_K$ is invertible since it is a diagonal matrix with no zero elements in its diagonal. Finally, $D_1 = 0$ guarantees that $\bbE_K^T\boldsymbol{\Psi}$ has rank $K$ whenever $K \leq P$ since it is a row-wise Vandermonde matrix with no repeated rows.
\end{myproof}

Condition~\emph{i)}, which is equivalent to that in Proposition~\ref{P:Recov_seeds_setup1}, guarantees that the low-pass filter with coefficients $\mathbf{h}^*$ does not eliminate any of the frequencies present in $\bby$. Condition~\emph{ii)} states requirements for recovery on both the seeding node and the global structure of the graph. The seeding node is required to be able to act on every active frequency ($U_1=0$), while the graph is required to have every active frequency distinguishable from each other ($D_1=0$). Condition~\emph{ii)} ensures that the rank of matrix $\bbE_K^T\diag(\widehat{\bbe}_1)\boldsymbol{\Psi}$ is equal to $K$ when $P \geq K$, guaranteeing that \eqref{eqn_solution_one_seed_low_pass} can be solved with respect to $\bbs_P$. For the particular case of $P=K$ the seeding values can be found as
\begin{equation}\label{eqn_proof_prop_seeds_setup3_010_solu}
\bbs_P= \left(\bbE_K^T\diag(\widehat{\bbe}_1)\boldsymbol{\Psi} \right)^{-1} \diag(\widehat{\mathbf{h}}_K^*)^{-1} \widehat{\bby}_K.
\end{equation}

When comparing the conditions~\emph{ii)} in Propositions~\ref{P:Recov_seeds_setup1} and~\ref{P:Recov_single_seeds_setup3}, we observe that for MN-ST seeding we should require a rank condition on a  submatrix of $\bbV^{-1}$. By contrast, for SN-MT seeding, the Vandermonde structure of $\boldsymbol{\Psi}$ allows reformulating the rank condition in terms of the graph related quantities $U_1$ and $D_1$, providing further insight on specifying the situations when recovery is possible. This dual behavior is also present when sampling graph signals. When following ``selection sampling'' \cite{SamplingKovacevic_without_Moura_15}, which is the counterpart of MN-ST interpolation, perfect reconstruction depends on the invertibility of a submatrix of $\bbV$, whereas when following an ``aggregation sampling'' scheme \cite{Oursampling_journal_2015}, which is the counterpart of SN-MT interpolation, the conditions for perfect reconstruction can be written in terms of specific graph related quantities.

Even though Proposition~\ref{P:Recov_single_seeds_setup3} guarantees perfect recovery under SN-MT seeding in a noiseless case, in noisy scenarios the selection of the seeding node is essential to reduce the reconstruction error. This is analyzed in Section~\ref{sec_imperfect_reconstruction} under a more general seeding scheme.

\subsection{Filter degree reduction in SN-MT seeding}\label{Ss:Single_seed_succesive_shift}

Mimicking the filter degree reduction technique presented in Section~\ref{Ss:low_pass_extrap_augmented}, SN-MT seeding can also implement a lower-degree filter if a higher number of seeding values is injected. To achieve this, we need the additional seeding values to generate a signal whose spectrum is zero for the inactive frequencies that are not eliminated by the filter. More specifically, the seeding values $\bbs_P$ and the filter coefficients $\bbh$ have to satisfy [cf.~\eqref{E:goal_freq_setup1_split_2_nonzeros_2}-\eqref{E:goal_freq_setup1_split_2_zeros_2}]
\begin{equation}\label{eqn_red_filt_SN-MT_seeding_010}
[\widehat{\bby}^T_K, \mathbf{0}^T_{P-K}]^T = \diag(\bbE_P^T \boldsymbol{\Psi}\mathbf{h} ) \bbE_P^T\diag(\widehat{\bbe}_1)\boldsymbol{\Psi}\bbs_P,
\end{equation}
\begin{equation}\label{eqn_red_filt_SN-MT_seeding_020}
\mathbf{0}_{N - P} = \diag(\bar{\bbE}_P^T \boldsymbol{\Psi}\mathbf{h})  \bar{\bbE}_P^T   \diag(\widehat{\bbe}_1)\boldsymbol{\Psi}\bbs_P,
\end{equation}
where $N-P$ is the number of frequency coefficients eliminated by the low-pass filter $\bbh$. As done in Section~\ref{Ss:low_pass_extrap_augmented}, $\bbh$ will be designed to solve \eqref{eqn_red_filt_SN-MT_seeding_020} for any choice of $\bbs_P$, while $\bbs_P$ will be chosen to solve the $P$ equations in \eqref{eqn_red_filt_SN-MT_seeding_010}. {A sufficient degree for $\bbh$ is presented next.}
{
%
\begin{myproposition}\label{prop_min_filter_SN-MT_seeding}
Let $U_2$ be the number of values in $\{[\widehat{\bbe}_1]_k\}_{k=K+1}^{N}$ that are zero and $D_2$ be the number of repeated values in $\{\lambda_k\}_{k\in \mathcal{K}_U}$, where $\mathcal{K}_U:=\{k\;\;|\;K<k\leq N\;\mathrm{and}\;[\widehat{\bbe}_1]_k\neq 0\}$. Then, \eqref{eqn_red_filt_SN-MT_seeding_020} can be solved with respect to $\bbh$ for any choice of $\bbs_P$ provided that {$L\!-\!1\geq \max(0, N\!-\!P\!-\!U_2\!-\!D_2$)}.
\end{myproposition}
\begin{myproof}
Notice that in \eqref{eqn_red_filt_SN-MT_seeding_020} the filter eliminates the \emph{last} $N-P$ frequencies, however, since the ordering is arbitrary, \emph{any} subset of $N-P$ frequencies (not containing the $K$ first ones) can be chosen to be annihilated by $\bbh$. Thus, our objective it to show that the proposed degree is enough to nullify a particular choice of $N-P$ frequency coefficients.
Define as $\ccalR$ the set of indices corresponding to zero elements in $\widehat{\bbe}_1$ or repeated rows in $\boldsymbol{\Psi}$. Since condition~\emph{ii)} in Proposition~\ref{P:Recov_single_seeds_setup3} must be satisfied -- otherwise, perfect recovery would be infeasible --, the cardinality of $\ccalR$ is $U_2+D_2$ and every index in $\ccalR$ must be greater than $K$.

{
First assume that $U_2 + D_2 < N - P$ and pick the $N-P$ frequencies to be eliminated to include the ones in $\ccalR$. This is equivalent to picking a frequency ordering such that every index in $\ccalR$ is greater than $P$. Based on $\ccalR$, define the selection matrices $\bbE_{\ccalR}:=[\bbe_{k_1}, \bbe_{k_2}, \ldots, \bbe_{k_{U_2+D_2}}]$ for all $k_i \in \ccalR$ and $\bar{\bbE}_{\ccalR} := [\bbe_{k_1}, \bbe_{k_2}, \ldots ,\bbe_{k_{N-P-U_2-D_2}}]$ for all $k_i \in \{P+1, \ldots, N\} \setminus \ccalR$ where $\setminus$ represents the set difference operator. Hence, the system of equations in \eqref{eqn_red_filt_SN-MT_seeding_020} can be split into two
\begin{align}
\mathbf{0}_{N-P-U_2-D_2} = \diag(\bar{\bbE}^T_{\ccalR} \boldsymbol{\Psi}\mathbf{h})  \bar{\bbE}^T_{\ccalR}   \diag(\widehat{\bbe}_1)\boldsymbol{\Psi}\bbs_P,
 \label{eqn_proof_prop_seeds_setup2_040}\\
\mathbf{0}_{U_2+D_2} = \diag(\bbE^T_{\ccalR} \boldsymbol{\Psi}\mathbf{h})  \bbE^T_{\ccalR}  \diag(\widehat{\bbe}_1)\boldsymbol{\Psi}\bbs_P.
\label{eqn_proof_prop_seeds_setup2_050}
\end{align}
Condition \eqref{eqn_proof_prop_seeds_setup2_040} can be guaranteed for any $\bbs_P$ if $\bbh=\bbh^*$, where $\bbh^*$ are the coefficients of a low-pass filter of degree $L\!-\!1=N\!-\!P\!-\!U_2\!-\!D_2$, as stated by the proposition. To complete the proof, we need to show that $\bbh=\bbh^*$ also guarantees that \eqref{eqn_proof_prop_seeds_setup2_050} holds. To see why this is the case, notice that $U_2$ rows of $ \bbE^T_{\ccalR}  \diag(\widehat{\bbe}_1)\boldsymbol{\Psi}$ are exactly zero, trivially satisfying \eqref{eqn_proof_prop_seeds_setup2_050} for any $\bbs_P$. Also, each of the remaining $D_2$ equations in \eqref{eqn_proof_prop_seeds_setup2_050} corresponds to a repeated eigenvalue and, thus, can be obtained by multiplying one of the $N-K-U_2-D_2$ homogenous equations in \eqref{eqn_proof_prop_seeds_setup2_040} and \eqref{eqn_red_filt_SN-MT_seeding_010} by a \emph{scalar}, guaranteeing that $\bbh^*$ also solves these $D_2$ equations.

For the case where $U_2 + D_2 \geq N - P$, we pick a frequency ordering such that every index greater than $P$ is contained in $\ccalR$. Thus, \eqref{eqn_red_filt_SN-MT_seeding_020} is implied by the homogenous equations in \eqref{eqn_red_filt_SN-MT_seeding_010}, and no filter (degree 0) is needed.}
\end{myproof}
}

Proposition~\ref{prop_min_filter_SN-MT_seeding} explicitly states that every additional seeding value decreases the required filter degree. However, in contrast to the situation for MN-ST, this reduction of the filter degree does not entail a reduction in the number of applications of the graph-shift operator, because it requires the length of the seeding phase to be extended. More interestingly, the additional seeding values can be used as a mean to guarantee perfect reconstruction when condition~\emph{i)} in Proposition~\ref{P:Recov_single_seeds_setup3} is not satisfied, as explained in Section~\ref{Ss:low_pass_extrap_augmented} for MN-ST seeding.

When $P\geq N\!-\!U_2\!-\!D_2$ the seeding phase suffices to recover the signal. This can be of interest in scenarios where the graph-shift operator $\bbS$ describes an intrinsic graph diffusion dynamic and the design of the filter coefficients is not feasible. It is also of interest if $\bby$ is not bandlimited. See Section~\ref{S:NumExper} for further discussions.

\subsection{Relation to classical interpolation}\label{Ss:Compar_with_low_pass_sinc_interp}

When $\bbS=\bbA_{dc}$, applying the $l$th power of $\bbS$ to a signal amounts to shifting the signal $l$ time instants. Consequently, the intermediate signal $\bbx$ obtained after the seeding phase in SN-MT reconstruction coincides with the seeding signal $\bbs$ in MN-ST reconstruction, provided that the seeding nodes are chosen adjacent to each other. Moreover, for the extreme case of the number of seeding values $P$ being enough to eliminate the filtering phase, which entails $L-1=0$, Proposition~\ref{prop_min_filter_SN-MT_seeding} requires setting $P=N$, because both $U_2$ and $D_2$ are zero if $\bbS=\bbA_{dc}$. The design of the $P=N$ seeding values $\bbs_P$ that guarantee that $\bbx=\bby$ can be carried out trivially by setting $\bbs_P=[x_1,\ldots,x_N]=\bby$.


\section{Multiple node - multiple time seeding}\label{S:shift_space_extrapolation}

In the more general multiple node - multiple time (MN-MT) seeding scheme, we can have several seeding signals ($\tau > 1$) \emph{and} we do not assume any structure on $\bbs^{(t)}$, so that any node may inject a seeding value at any given time. We concatenate the $\tau$ seeding signals into the $N\tau \times 1$ vector $\underline{\bbs}$ defined as $\underline{\bbs} := \mathrm{vec}([\bbs^{(\tau-1)}, \bbs^{(\tau-2)}, \ldots, \bbs^{(0)}]^T).$

Defining the $N \times N^2$ matrix $\boldsymbol{\Theta} := [\diag(\widehat{\bbe}_1), \ldots, \diag(\widehat{\bbe}_N)]$, we may relate $\bbx$ to $\underline{\bbs}$ as stated in the following lemma.
%
\begin{mylemma}\label{lem_seeding_signal_MN-MT}
The frequency representation of the intermediate signal $\bbx$ in MN-MT seeding is given by
\begin{equation}\label{eqn_freq_output_extension_matrix_MN-MT}
\widehat{\bbx} = \boldsymbol{\Theta} (\bbI \otimes \boldsymbol{\Psi}) \underline{\bbs},
\end{equation}
where $\otimes$ represents the Kronecker product.
\end{mylemma}
\begin{myproof}
If we denote by $\bbx := \bbx^{(\tau-1)}$ the signal obtained after the seeding phase, it holds that [cf. \eqref{eqn_signal_y_time_t}]
\begin{equation}\label{eqn_percolated_tau_general}
\bbx = \sum_{l=0}^{\tau-1} \bbS^l \bbs^{(\tau-1-l)} = \sum_{l=0}^{\tau-1} \bbS^l \left( \sum_{i=1}^N s_i^{(\tau-1-l)} \bbe_i \right) = \sum_{i=1}^N \bbH_i \bbe_i,
\end{equation}
where the filter $\bbH_i$ is given by
\begin{equation}\label{eqn_def_filter_bbH_i}
\bbH_i = \sum_{l=0}^{\tau-1} s_i^{(\tau-1-l)} \bbS^l.
\end{equation}
Writing the input-output relationship of those filters in the frequency domain, we have that [cf. \eqref{E:Filter_input_output_freq}]
\begin{equation}\label{eqn_freq_output_extension}
\widehat{\bbx} = \sum_{i=1}^N \diag(\boldsymbol{\Psi} \bbs_i) \widehat{\bbe}_i = \sum_{i=1}^N \diag(\widehat{\bbe}_i) \boldsymbol{\Psi} \bbs_i.
\end{equation}
Recalling the definitions of $\boldsymbol{\Theta}$ and $\underline{\bbs}$, the sum in \eqref{eqn_freq_output_extension} can be written in matrix form, giving rise to \eqref{eqn_freq_output_extension_matrix_MN-MT}.
\end{myproof}

As was the case for Lemma~\ref{lem_seeding_phase} in SN-MT seeding, Lemma~\ref{lem_seeding_signal_MN-MT} leverages the reinterpretation of the seeding phase as the application of a filter -- in this case, $N$ different filters, one per node -- to the canonical basis vectors [cf. \eqref{eqn_percolated_tau_general}]. The coefficients of the filter associated with the $i$-th node are given by the values injected by that $i$-th node [cf. \eqref{eqn_def_filter_bbH_i}].
Notice that, as expected, \eqref{eqn_freq_output_extension_matrix_MN-MT} reduces to \eqref{E:interp_shift_ver3_2} whenever the seeding values are forced to be zero for every seeding node except for the first one.

To analyze the output of the filtering phase $\bbz$, recall that $\widehat{\mathbf{h}}_K^*=\bbE_K^T\boldsymbol{\Psi}\mathbf{h}^*$ denotes the response of a low-pass filter in the active frequencies. Mimicking the procedure in Section~\ref{S:Extrapolating_GS_succesive}, we find that the active frequency coefficients in $\bby$ can be written in terms of the seeding values $\underline{\bbs}$ as
\begin{equation}\label{eqn_extension_low_pass}
\widehat{\bby}_K = \widehat{\bbz}_K = \diag(\widehat{\mathbf{h}}_K^*) \bbE_K^T \boldsymbol{\Theta} (\bbI \otimes \boldsymbol{\Psi}) \underline{\bbs}.
\end{equation}
The system of equations in \eqref{eqn_extension_low_pass} is underdetermined, since the $K$ values in $\widehat{\bby}_K$ can be reconstructed using the $N\tau$ values in $\underline{\bbs}$. However, our focus is on the case where only $P\ll N\tau$ seeding values are injected during the seeding phase. To this extent, we introduce the $P \times N\tau$ \emph{selection} matrix $\bbC$ whose elements are binary $C_{ij}\in\{0,1\}$ and satisfy $\sum_j C_{ij}=1$ and $\sum_i C_{ij}\leq 1$ for all $i$ and $j$, respectively. Since the matrix has exactly one 1 in every row, $\bbC$ selects $P$ seeding values among the $N\tau$ node-time pairs. If we denote by $\underline{\bbs}_P:= \bbC \underline{\bbs}$ the vector containing these $P$ seeding values, \eqref{eqn_extension_low_pass} can be rewritten as [cf. \eqref{E:goal_freq_setup1_split_2_nonzeros_after_filter} and \eqref{eqn_solution_one_seed_low_pass}]
\begin{equation}\label{eqn_extension_low_pass_2}
\widehat{\bby}_K = \diag(\widehat{\mathbf{h}}_K^*) \bbE_K^T \boldsymbol{\Theta} (\bbI \otimes \boldsymbol{\Psi}) \bbC^T \underline{\bbs}_P.
\end{equation}
To resemble the structure of previous sections, the conditions under which \eqref{eqn_extension_low_pass_2} can be solved with respect to $\underline{\bbs}_P$ are given in the form of a proposition.
%
\begin{myproposition}\label{P:Recov_seeds_setup_general}
The system of $K$ equations in \eqref{eqn_extension_low_pass_2} is guaranteed to have a solution with respect to $\underline{\bbs}_P$ if the following two conditions hold:\\
 i) $\lambda_{k_1}\neq\lambda_{k_2}$ for all $(\lambda_{k_1},\lambda_{k_2})$ such that $k_1 \leq K$ and $k_2>K$, \\
ii) $\mathrm{rank}(\bbE_K^T \boldsymbol{\Theta} (\bbI \otimes \boldsymbol{\Psi}) \bbC^T)\geq K$.
\end{myproposition}
\begin{myproof}
Condition~\emph{i)} is required to guarantee that $\diag(\widehat{\mathbf{h}}_K^*)$ is invertible (cf. proof of Proposition~\ref{P:Recov_seeds_setup1}). This allows us to rewrite \eqref{eqn_extension_low_pass_2} as
\begin{equation}\label{eqn_extension_low_pass_2_proof_010}
\diag(\widehat{\mathbf{h}}_K^*)^{-1} \widehat{\bby}_K =  \bbE_K^T \boldsymbol{\Theta} (\bbI \otimes \boldsymbol{\Psi}) \bbC^T \underline{\bbs}_P.
\end{equation}
To guarantee that the system of equations in \eqref{eqn_extension_low_pass_2_proof_010} has at least one solution, we need condition~\emph{ii)} to hold.
\end{myproof}

Condition~\emph{i)}, also present in Propositions~\ref{P:Recov_seeds_setup1} and~\ref{P:Recov_single_seeds_setup3}, guarantees that the filtering phase does not annihilate any of the frequencies present in $\bby$. Condition~\emph{ii)} requires, at the very least, $P\geq K$. However, there may be cases where setting $P=K$ can fail as stated in the discussion ensuing Proposition~\ref{P:Recov_seeds_setup1}.

Mimicking the developments in Sections~\ref{Ss:low_pass_extrap_augmented} and~\ref{Ss:Single_seed_succesive_shift}, for the general case of MN-MT seeding, additional seeding values can be used to reduce the degree of the low-pass filter. Indeed, for every extra seeding value the degree of the filter needed decreases by one, reducing the communication cost of the reconstruction scheme. Moreover, these extra seeding values can be used to obtain perfect reconstruction even when condition~\emph{i)} in Proposition~\ref{P:Recov_seeds_setup_general} is violated, as explained in Section~\ref{Ss:low_pass_extrap_augmented}.

The selection matrix $\bbC$ can be designed so that condition~\emph{ii)} in Proposition~\ref{P:Recov_seeds_setup_general} is satisfied, guaranteeing perfect recovery. Furthermore, for the cases in which perfect reconstruction is infeasible due to, e.g., the presence of noise, the choice of $\bbC$ can be optimized to achieve robust recovery, as analyzed in the following section.

\begin{remark}\label{rem_MN-MT_as_generalization}\normalfont
When $\bbC$ in \eqref{eqn_extension_low_pass_2} selects the first $P$ elements of $\underline{\bbs}$, \eqref{eqn_extension_low_pass_2} reduces to \eqref{eqn_solution_one_seed_low_pass} and SN-MT reconstruction is recovered. Similarly, if $\bbC$ selects the elements of $\underline{\bbs}$ in positions $1, \tau+1, \ldots, P\tau+1$, then \eqref{eqn_extension_low_pass_2} reduces to \eqref{E:goal_freq_setup1_split_2_nonzeros_after_filter} as in MN-ST reconstruction.
\end{remark}

\section{Imperfect reconstruction}\label{sec_imperfect_reconstruction}

We study two settings where perfect reconstruction is infeasible: insufficient number of seeding values (Section~\ref{sec_insufficient_seeding_values}) and additive noise in the injections (Section~\ref{sec_noise_when_injecting}). The analysis is focused on the MN-MT seeding scheme, since the results for MN-ST and SN-MT can be obtained by particularizing the value of the selection matrix $\bbC$ (cf. Remark~\ref{rem_MN-MT_as_generalization}).

\subsection{Insufficient seeding values}\label{sec_insufficient_seeding_values}

When the number of seeding values $P$ is not enough to achieve perfect reconstruction, the goal is to minimize a pre-specified error metric between the reconstructed signal $\bbz$ and the original $K$-bandlimited graph signal $\bby$. Three different design scenarios are considered. In the first one, the seeding values $\underline{\bbs}_P$ are designed assuming that both $\bbh$ and $\bbC$ are fixed. The second scenario addresses the joint design of $\underline{\bbs}_P$ and $\bbh$. In the last one, the joint design of $\underline{\bbs}_P$ and $\bbC$ is performed.

\subsubsection{Designing the seeding values $\underline{\bbs}_P$}

Assume that condition~\emph{i)} in Proposition~\ref{P:Recov_seeds_setup_general} holds and recall that $\mathbf{h}^*$ stands for the coefficients of a low-pass filter that eliminates all the frequencies $k>K$.
Then, the first $K$ frequency coefficients $\widehat{\bbz}_K$ of the reconstructed signal $\bbz$ are obtained as [cf. \eqref{eqn_extension_low_pass_2}] %
\begin{equation}\label{E:approx_signal_reconst_after_filter}
\widehat{\bbz}_K= \diag(\widehat{\mathbf{h}}_K^*) \bbE_K^T \boldsymbol{\Theta} (\bbI \otimes \boldsymbol{\Psi}) \bbC^T \underline{\bbs}_P.
\end{equation}
Since we assume insufficient seeding values, i.e., $P < K$, obtaining $\widehat{\bby}_K=\widehat{\bbz}_K$ is in general infeasible. A reasonable approach is to design $\underline{\bbs}_P$ to minimize the energy of the reconstruction error. Defining the matrix
\begin{equation}\label{E:imperfect_reconstruction_err_cov_matrix_v00}
\boldsymbol{\Phi}_{K}:= \diag(\widehat{\mathbf{h}}_K^*) \bbE_K^T \boldsymbol{\Theta} (\bbI \otimes \boldsymbol{\Psi}),
\end{equation}
the optimal seeding values $\underline{\bbs}_P^*$ can be obtained as
\begin{equation}\label{E:imperfect_reconstruction_known_support_prob_formul}
\underline{\bbs}_P^*:=\arg\min_{\underline{\bbs}_P}\|\bby-\bbV_K \boldsymbol{\Phi}_{K} \bbC^T \, \underline{\bbs}_P\|_2^2,
\end{equation}
where, we recall, $\bbV_K := \bbV \bbE_K$. The minimization problem in \eqref{E:imperfect_reconstruction_known_support_prob_formul} has the well-known closed-form solution \cite{KayBook}
\begin{align}
\underline{\bbs}_P^* = (\bbC\boldsymbol{\Phi}_{K}^H\, \boldsymbol{\Phi}_{K}\bbC^T)^{-1} \bbC\boldsymbol{\Phi}_{K}^H\, \bbV_K^H \bby, \label{E:imperfect_reconstruction_known_support_closed_form_sol}
\end{align}
where we assume that the \emph{fixed} seeding locations $\bbC$ lead to a matrix $\boldsymbol{\Phi}_{K}\bbC^T$ that has full column rank.
%
%
With $\bbeps:=\bby \!- \!\bbz$ denoting the reconstruction error, its energy can be written as
\begin{align}
\!\| \bbeps \|_2^2 \!= \! \bby^H \bbV_K \! \left( \! \bbI - \boldsymbol{\Phi}_{K} \bbC^T (\bbC\boldsymbol{\Phi}_{K}^H\, \boldsymbol{\Phi}_{K}\bbC^T)^{\!-1} \bbC \boldsymbol{\Phi}_{K}^H \! \right) \! \bbV^H_K \bby.
\label{E:imperfect_reconstruction_known_support_closed_form_sol_3}
\end{align}
Notice that, since $\bbh^*$ is given, the reconstruction error is zero for the frequency components $k > K$.

\subsubsection{Designing the seeding values $\underline{\bbs}_P$ and the filter coefficients $\bbh$}\label{sub_sub_designing_seeding_filter}

When perfect reconstruction is infeasible, carrying out a separate optimization of $\bbh$ and $\underline{\bbs}_P$, where $\bbh$ is designed to filter the frequencies not present in $\bby$ and $\underline{\bbs}_P$ is designed to match the spectrum of $\bby$ in the active frequencies, is not jointly optimal. Minimization of the reconstruction error by jointly designing $\underline{\bbs}_P$ and $\bbh$ is briefly discussed next. Notice that the $N$ frequency coefficients -- as opposed to just the first $K$ coefficients -- of the reconstructed signal $\widehat{\bbz}$ are [cf. \eqref{eqn_extension_low_pass_2}]
\begin{equation}\label{E:imperfect_reconstruction_known_support_prob_formul_1_010}
\widehat{\bbz} = \diag(\boldsymbol{\Psi}\bbh) \bbE_K^T \boldsymbol{\Theta} (\bbI \otimes \boldsymbol{\Psi}) \bbC^T \underline{\bbs}_P.
\end{equation}
Hence, if the objective is to minimize $\|\bbeps\|_2^2$, we have that
\begin{align}\label{E:imperfect_reconstruction_known_support_prob_formul_1_020}
\{\underline{\bbs}_P^*,\,& \bbh^* \}:= \argmin_{\{\underline{\bbs}_P, \bbh\}} \| \bby - \bbV \, \widehat{\bbz} \|_2^2 \\
&= \argmin_{\{\underline{\bbs}_P, \bbh\}}\|\bby- \bbV \diag(\boldsymbol{\Psi}\bbh) \bbE_K^T \boldsymbol{\Theta} (\bbI \otimes \boldsymbol{\Psi}) \bbC^T \, \underline{\bbs}_P\|_2^2, \nonumber
\end{align}
which is a \emph{bilinear} optimization. Bilinear problems are non-convex, but there is a large amount of works dealing with their analysis and efficient solution \cite{McCormick_76_bilinear,Konno_76_bilinear,choudhary2014identifiability}.

The formulation in \eqref{E:imperfect_reconstruction_known_support_prob_formul_1_020} considers that $\bbh^*$ can be chosen as a function of the signal to reconstruct $\bby$. In applications where this is not feasible, formulating the optimal design requires additional assumptions on $\bby$. If the distribution of $\bby$ is known, a two-stage stochastic programming approach can be pursued \cite{shapiro2014lectures}. In the second stage, $\bbh$ in \eqref{E:imperfect_reconstruction_known_support_prob_formul_1_020} is considered given and the optimal $\underline{\bbs}^*_P$ is obtained as the minimizer of $\|\bbeps(\bbh,\bby,\underline{\bbs}_P)\|_2^2$, which is a function of $\bbh$ and $\bby$ [cf. \eqref{E:imperfect_reconstruction_known_support_closed_form_sol}]. In the first stage, the solution of the second stage $\underline{\bbs}^*_P(\bbh,\bby)$ and the distribution of $\bby$ are leveraged to write the expectation of the reconstruction error in \eqref{E:imperfect_reconstruction_known_support_prob_formul_1_020} as $\bar{\epsilon}(\bbh):=\mathop{{}\mathbb{E}}_{\bby}[\|\bbeps(\bbh,\bby,\underline{\bbs}^*_P(\bbh,\bby))\|_2^2]$, which only depends on $\bbh$. The optimum $\bbh^*$ is then the minimizer of the expected error $\bar{\epsilon}(\bbh)$.
{Notice that this two-stage approach is used in Sections~\ref{S:Extrapolating_GS_subset_nodes}, \ref{S:Extrapolating_GS_succesive}, and \ref{S:shift_space_extrapolation} to find conditions for perfect recovery, where bandlimitedness is the prior knowledge of $\bby$.
}

\subsubsection{Designing the seeding values $\underline{\bbs}_P$ and the seeding locations $\bbC$}\label{subsubsec_designing_seeding_values_seeding_locations}

Suppose now that one can select the specific nodes and time instants where the injections take place. This amounts to choosing the $P$ entries of $\bbC$ that are non-zero, which is a combinatorial problem. Although for small networks one could try all possible choices of $\bbC$ and select the one leading to the smallest reconstruction error, for general networks a more scalable approach is required. To formulate the corresponding optimization problem $\underline{\bbs}_P=\bbC\, \underline{\bbs}$ is substituted into \eqref{E:imperfect_reconstruction_known_support_prob_formul}. After that, the product $\bbC^T \bbC$ is rewritten as $\diag(\bbc)$ where $\bbc$ is a binary selection vector of dimension $N\tau \times 1$. Note that having $c_i=1$ indicates that at time $t=\mathrm{mod}_\tau(N\tau-i)$ the node $(i+t)/\tau$ injects a seeding value. With this notation, the joint design of $\underline{\bbs}_P$ and $\bbc$ amounts to solving
\begin{align}\label{E:imperfect_reconstruction_unknown_support}
\{\underline{\bbs}^*, \bbc^*\}&:=\argmin_{\{\underline{\bbs}, \bbc\}}\|\bby-\bbV_K \boldsymbol{\Phi}_{K} \diag(\bbc) \, \underline{\bbs}\|_2^2+\gamma\| \bbc\|_0 \nonumber \\
&\text{s.t.} \qquad \bbc \in \{0, 1\}^{N\tau},
\end{align}
where $\boldsymbol{\Phi}_{K}$ is defined in \eqref{E:imperfect_reconstruction_err_cov_matrix_v00}. In \eqref{E:imperfect_reconstruction_unknown_support} each seeding location used is penalized with a constant cost $\gamma$. By tuning $\gamma$, the desired level of sparsity of $\bbc$ can be achieved. {Problem \eqref{E:imperfect_reconstruction_unknown_support} can be further simplified by setting $\bbd : = \diag(\bbc) \, \underline{\bbs}$ and requiring sparsity on $\bbd$
\begin{align}\label{E:imperfect_reconstruction_unknown_support_2}
\bbd^*&:=\argmin_\bbd \|\bby-\bbV_K \boldsymbol{\Phi}_{K} \bbd \|_2^2+\gamma\| \bbd\|_0.
\end{align}
Among other advantages, the formulation in \eqref{E:imperfect_reconstruction_unknown_support_2} is amenable to relaxations that reduce the computational complexity required to find a solution. A straightforward approach is to relax the problem by replacing the 0-norm with the 1-norm to obtain a convex formulation.}

As in problem \eqref{E:imperfect_reconstruction_known_support_prob_formul_1_020}, the design in \eqref{E:imperfect_reconstruction_unknown_support} and its subsequent simplification in \eqref{E:imperfect_reconstruction_unknown_support_2} assume that the seeding nodes can be chosen as a function of $\bby$. For applications where this is not convenient, a two-stage stochastic programming approach similar to the one described for \eqref{E:imperfect_reconstruction_known_support_prob_formul_1_020} can also be used in solving \eqref{E:imperfect_reconstruction_unknown_support}. Last but not least, although computationally challenging, a joint optimization of $\underline{\bbs}_P$, $\bbh$ and $\bbC$ can be pursued by combining the approaches in Sections~\ref{sub_sub_designing_seeding_filter}~and~\ref{subsubsec_designing_seeding_values_seeding_locations}.

\subsection{Noise when injecting the seeding values}\label{sec_noise_when_injecting}

The conditions for perfect reconstruction stated in Propositions~\ref{P:Recov_seeds_setup1},~\ref{P:Recov_single_seeds_setup3} and~\ref{P:Recov_seeds_setup_general} require the seeding values to be the exact solution of \eqref{E:goal_freq_setup1_split_2_nonzeros_after_filter}, \eqref{eqn_solution_one_seed_low_pass}, and \eqref{eqn_extension_low_pass_2}, respectively. However, in real applications, the injected values can be corrupted with additive noise. This noise can be either attenuated or amplified when the signal percolates through the graph via the successive applications of $\bbS$. The goal of this section is to quantify the reconstruction error and to discuss seeding selection schemes tailored to these operating conditions. Their performance will be illustrated via numerical simulations in Section~\ref{S:NumExper}.

Let us assume that the injected signal is $\underline{\bbs}_P + \bbw_P$, where $\bbw_P$ is a $P \times 1$ noise vector with zero mean and covariance $\bbR_{\bbw}$. The active frequencies of the reconstructed signal can then be written as $\widehat{\bbz}_K = \boldsymbol{\Phi}_{K} \bbC^T (\underline{\bbs}_P + \bbw_P)$ [cf.~\eqref{eqn_extension_low_pass_2}~and~\eqref{E:imperfect_reconstruction_err_cov_matrix_v00}]. From this, we may obtain the reconstruction error as
\begin{equation}\label{eqn_noise_injected_values_020}
\bbeps = \bbV_K(\widehat{\bbz}_K - \widehat{\bby}_K) = \bbV_K \boldsymbol{\Phi}_{K} \bbC^T \bbw_P,
\end{equation}
with covariance matrix
\begin{equation}\label{eqn_noise_injected_values_030}
\bbR_{\bbeps} = \mathop{{}\mathbb{E}}(\bbeps \bbeps^H) =\bbV_K \boldsymbol{\Phi}_{K} \bbC^T \bbR_{\bbw} \bbC \boldsymbol{\Phi}^H _{K} \bbV^H_K.
\end{equation}
Ideally, $\bbC$ should be designed to select the seeding nodes and time instants that minimize the reconstruction error, which can be quantified as a function of $\bbR_{\bbeps}$. In what follows, we will focus on minimizing the mean squared error (MSE), which is achieved by minimizing $\trace(\bbR_{\bbeps})$. However, similar approaches can be followed to minimize other commonly used error metrics such as $\lambda_{\max}(\bbR_{\bbeps})$ and $\log(\det(\bbR_{\bbeps}))$ \cite{pukelsheim1993optimal}.

To illustrate the design of $\bbC$, two particular scenarios are considered.
In the first one, we assume i) that the frequency coefficients $\widehat{\bby}_K$ are zero mean with covariance $\bbR_{\widehat{\bby}} = \bbI$ and ii) that the noise $\bbw_P$ is zero mean with covariance $\bbR_{\bbw} = \sigma^2 \mathop{{}\mathbb{E}}(\|\underline{\bbs}_P\|_2^2) \bbI$. Note that assumption ii) is meaningful if the system operates under a constant signal-to-noise ratio (SNR) regime. In the second scenario, we assume the noise is also uncorrelated but its power is independent of that of the seeding signal, so that $\bbR_{\bbw} = \sigma^2 \bbI$.

\begin{figure*}[t]
\centering

\begin{subfigure}{.37\textwidth}
  \centering
  \input{figures/synthetic_graph_and_signals.tex}
  \caption{}
  \label{fig:sub1_synthetic}
\end{subfigure}%
\begin{subfigure}{.31\textwidth}
  \centering
  \includegraphics[width=0.9\textwidth]{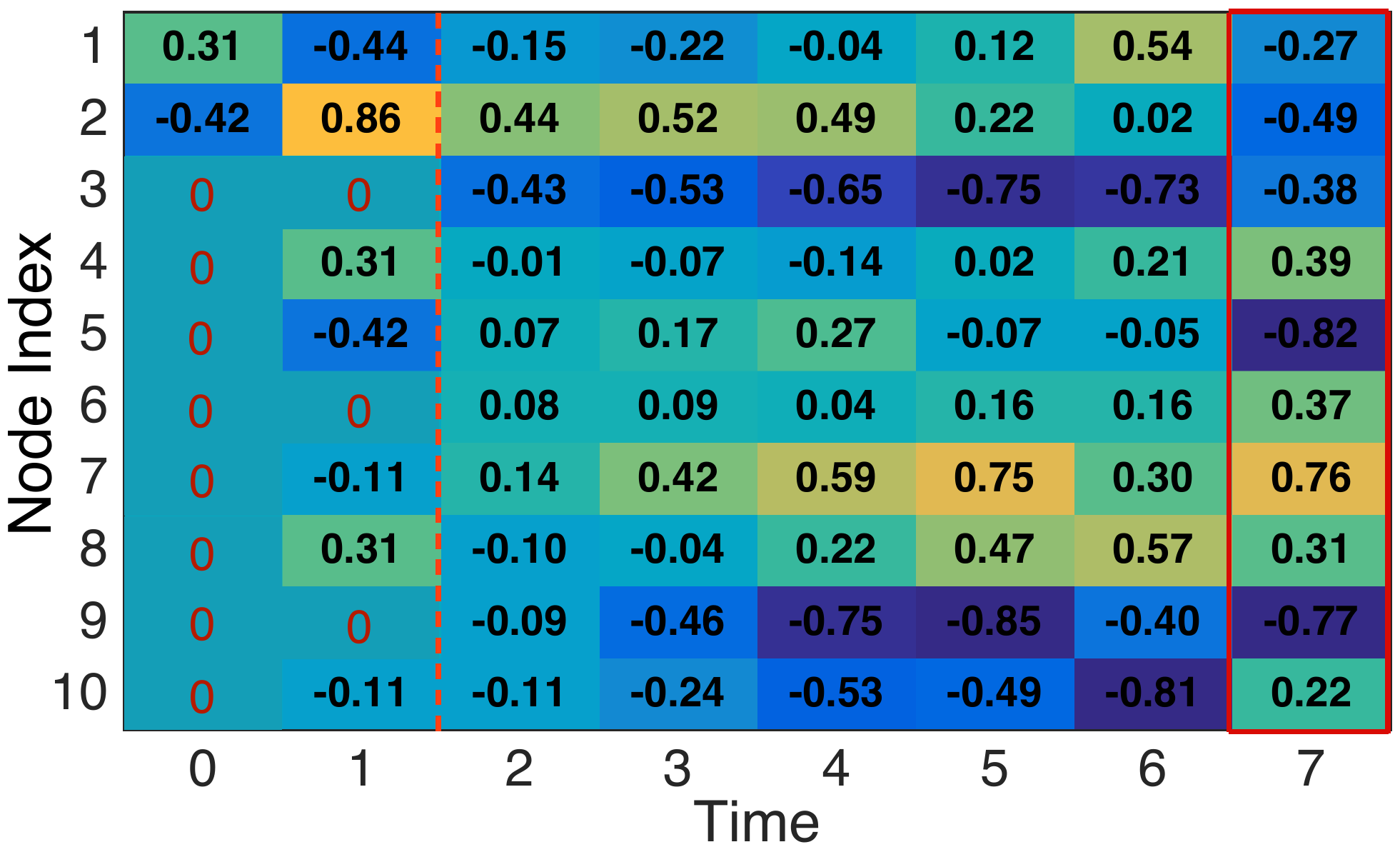}
  \caption{}
  \label{fig:sub2_synthetic}
\end{subfigure}%
\begin{subfigure}{.31\textwidth}
  \centering
  \includegraphics[width=0.9\textwidth]{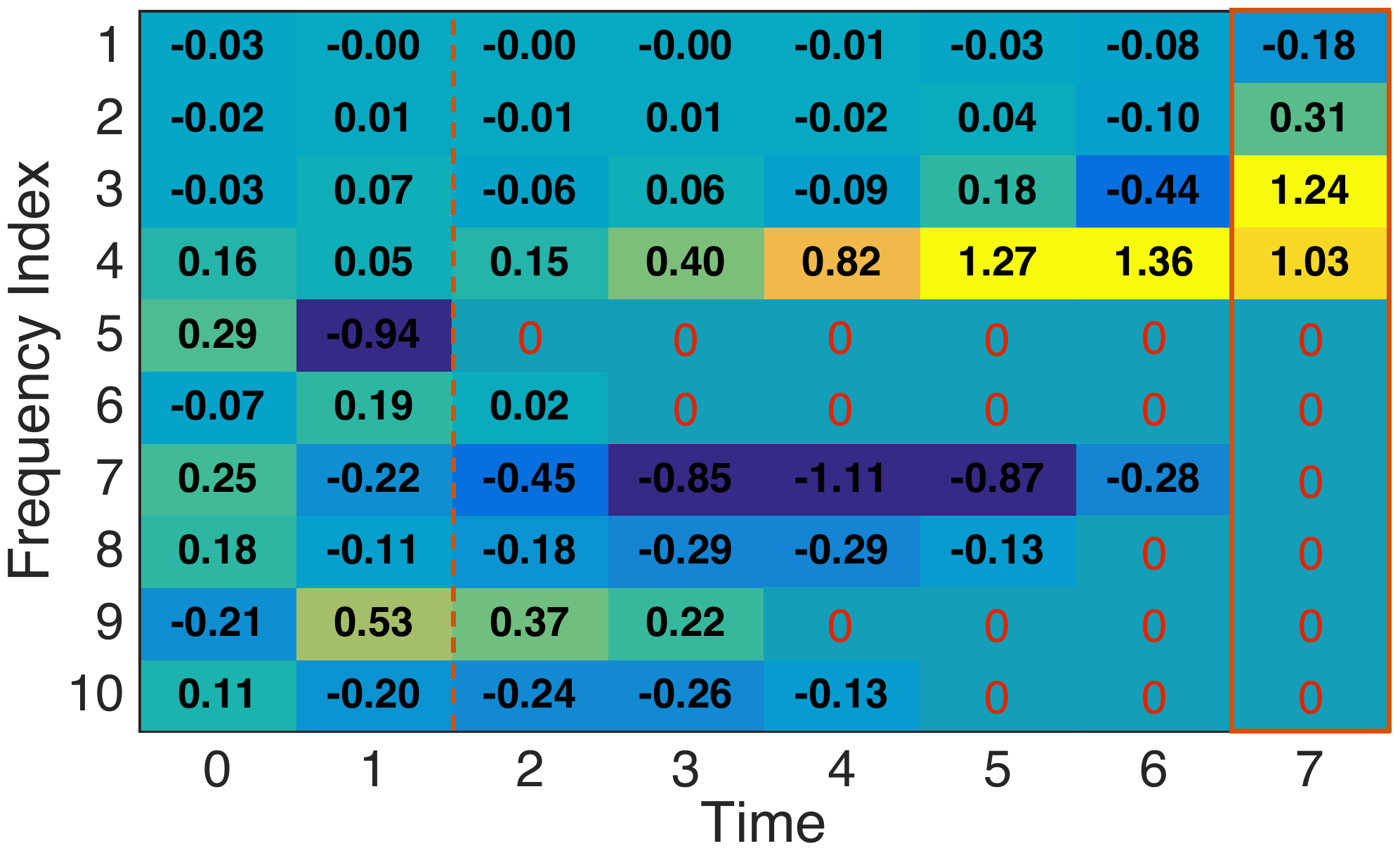}
  \caption{}
  \label{fig:sub3_synthetic}
\end{subfigure}
\vspace{-0.05in}
\caption{Perfect recovery of a bandlimited graph signal. (a) The graph $\ccalG$, the target signal to recover $\bby$ and its frequency representation $\widehat{\bby}$. (b) Evolution of the reconstructed signal. The seeding and filtering phases are separated by a dotted line and the recovered signal is framed in red. (c) Evolution of the frequency components of the reconstructed signal. Successive annihilation during the filtering phase is observed.}
\label{fig_synthetic_experiments}
\vspace{-0.1in}
\end{figure*}

For convenience, the optimal seeding strategy for the first scenario is presented in the form of a lemma.
%
\begin{mylemma}
Suppose that $\widehat{\bby}_K$ and $\bbw_P$ are drawn from zero-mean distributions with covariances $\bbR_{\widehat{\bby}} = \bbI$ and $\bbR_{\bbw} = \sigma^2\mathop{{}\mathbb{E}}(\|\underline{\bbs}_P\|_2^2) \bbI$, respectively. Then, the selection $\bbc^*$ that minimizes the MSE of the reconstruction is given by
\begin{align}\label{eqn_noise_injected_values_060}
\bbc^*& \!\! := \! \argmin_{\bbc} \, \trace \!\Big( \! \big( \! \boldsymbol{\Phi} _{K} \diag(\bbc) \boldsymbol{\Phi}^H _{K}\! \big)^{\!-1} \Big)  \trace  \Big( \! \boldsymbol{\Phi} _{K} \diag(\bbc) \boldsymbol{\Phi}^H _{K}\! \Big) \nonumber \\
&\qquad\,\,  \text{s.t.} \qquad \bbc \in \{0, 1\}^{N\tau}, \quad \| \bbc \|_0 = P.
\end{align}
\end{mylemma}
\begin{myproof}
To prove the lemma, we need to show that the minimization of the objective in \eqref{eqn_noise_injected_values_060} is equivalent to the minimization of $\trace(\bbR_{\bbeps})$. By substituting $\bbR_{\bbw} = \sigma^2 \mathop{{}\mathbb{E}}(\|\underline{\bbs}_P\|_2^2) \bbI$ and $\diag(\bbc):=\bbC^T \bbC$ into \eqref{eqn_noise_injected_values_030}, it follows that
\begin{equation}\label{eqn_noise_injected_values_070}
\trace(\bbR_{\bbeps}) = \sigma^2\mathop{{}\mathbb{E}}(\|\underline{\bbs}_P\|_2^2) \trace(\bbV_K \boldsymbol{\Phi}_{K} \diag(\bbc) \boldsymbol{\Phi}^H _{K} \bbV^H_K).
\end{equation}
Since the trace is invariant to cyclic permutations and $\bbV_K^H \bbV_K = \bbI$, we have that
\begin{equation}\label{eqn_noise_injected_values_080}
\trace(\bbR_{\bbeps}) = \sigma^2\mathop{{}\mathbb{E}}(\|\underline{\bbs}_P\|_2^2) \trace(\boldsymbol{\Phi}_{K} \diag(\bbc) \boldsymbol{\Phi}^H _{K}).
\end{equation}
To find an expression for $\mathop{{}\mathbb{E}}(\|\underline{\bbs}_P\|_2^2)$, we leverage the fact that $\widehat{\bby}_K = \boldsymbol{\Phi}_{K} \bbC^T \underline{\bbs}_P$ [cf.~\eqref{eqn_extension_low_pass_2}~and~\eqref{E:imperfect_reconstruction_err_cov_matrix_v00}] to write $\underline{\bbs}_P =  \bbC \boldsymbol{\Phi}^H_{K} (\boldsymbol{\Phi}_{K} \bbC^T \bbC \boldsymbol{\Phi}^H_{K})^{-1} \widehat{\bby}_K$ and, consequently, to write $\|\underline{\bbs}_P\|_2^2$ as
\begin{equation}\label{eqn_noise_injected_values_090}
\|\underline{\bbs}_P\|_2^2 = \underline{\bbs}^H_P \underline{\bbs}_P = \widehat{\bby}^H_K (\boldsymbol{\Phi}_{K} \diag(\bbc) \boldsymbol{\Phi}^H_{K})^{-1} \widehat{\bby}_K.
\end{equation}
Using the expression for the expected value of a quadratic form, it follows that
\begin{equation}\label{eqn_noise_injected_values_100}
 \mathop{{}\mathbb{E}}(\|\underline{\bbs}_P\|_2^2) = \trace \!\Big( \! \big( \boldsymbol{\Phi} _{K} \diag(\bbc) \boldsymbol{\Phi}^H _{K} \big)^{\!-1} \Big).
\end{equation}
Upon replacing \eqref{eqn_noise_injected_values_100} into \eqref{eqn_noise_injected_values_080} and recalling that $\sigma^2$ does not depend on $\bbc$, the expression in \eqref{eqn_noise_injected_values_060} follows.
\end{myproof}

\noindent The statistical assumption on $\widehat{\bby}_K$ allows us to design $\bbc^*$ such that the \emph{expected} performance of the reconstruction scheme is optimized. In this way, the choice of the seeding nodes and instants is independent of the particular signal being reconstructed.

Even though obtaining general relaxations to efficiently approximate the non-convex problem in \eqref{eqn_noise_injected_values_060} is out of the scope of the paper, we can gain intuition by specializing \eqref{eqn_noise_injected_values_060} for time-varying signals, i.e., by setting $\bbS = \bbA_{dc}$. For SN-MT seeding, where designing $\bbc$ boils down to selecting the seeding node, it can be shown that the objective in \eqref{eqn_noise_injected_values_060} does not depend on the particular node chosen. This is as it should be, since in the directed cycle every node is topologically indistinguishable from the others. For MN-ST seeding, the best strategy is to uniformly distribute the seeding nodes, as we formally state next.
%
\begin{myproposition}
Suppose that the problem in \eqref{eqn_noise_injected_values_060} is particularized for the case of MN-ST seeding of time-varying signals using an ideal low-pass filter. Then, if $K=P=N/\theta$, it holds that the optimal seeding strategy selects the nodes in positions $1, 1+\theta, \ldots, 1+(K-1)\theta$.
\end{myproposition}
\begin{myproof}
When $\bbS=\bbA_{dc}$ we have that: a) $\bbV = \bbF$; b) $\widehat{\bbh}^*_K = \alpha \mathbf{1}_K$ for some constant $\alpha$ where $\mathbf{1}_K$ is the $K \times 1$ vector of all ones -- since we are considering an ideal low-pass filter --; and c) $\bbc$ only can take nonzero values in positions $i=1, 1+ \tau, \ldots, 1+ (N-1) \tau$ (since we are considering MN-ST seeding). Leveraging a), b) and c), problem \eqref{eqn_noise_injected_values_060} can be reformulated as
\begin{align}\label{eqn_noise_injected_values_110}
\bbc^*& := \argmin_{\bbc} \, \trace \left( \bbM^{-1} \right) \trace \left(  \bbM \right) \\
& \text{s.t.} \,\,  \bbM = \bbE_K^T \bbF^H \diag(\bbc) \bbF \bbE_K, \,\,\, \bbc \in \{0, 1\}^{N}, \, \| \bbc \|_0 = K, \nonumber
\end{align}
where $\bbc$ selects $K$ seeding nodes out of the $N$ possible ones. First, notice that $\trace(\bbM)$ does not depend on the particular choice of $\bbc$. To see why this is true, we denote by $\ccalI(\bbc)$ the set containing the indices of the $K$ seeding nodes selected by $\bbc$. Then, we can exploit the structure in $\bbF$ to write
\begin{align}\label{eqn_noise_injected_values_120}
\trace(\bbM) = \sum_{i \in \ccalI(\bbc)} \sum_{j=0}^{K-1} \left| \frac{1}{\sqrt{N}}e^{+\mathfrak{j}\frac{2\pi}{N}(i-1)(j-1)} \right|^2 = \frac{K^2}{N},
\end{align}
which does not depend on $\bbc$. Hence, the optimal $\bbc^*$ in \eqref{eqn_noise_injected_values_110} can be found as the one minimizing $\trace \left( \bbM^{-1} \right)$.

If we denote by $\{\gamma_i\}_{i=1}^K$ the $K$ eigenvalues of $\bbM$, our goal is then to find the $\bbc^*$ that minimizes $\sum_i 1/\gamma_i$. Given that all $\gamma_i$ are nonnegative ($\bbM$ is positive semi-definite) and \eqref{eqn_noise_injected_values_120} implies that $\sum_i \gamma_i = K^2/N$, the minimization is achieved by setting $\gamma_1 = \gamma_2 = \ldots = \gamma_K = K/N$. Hence, if we show that uniform seeding leads to $\gamma_i=K/N$ for all $i$, the proof concludes.
To show this, notice that under uniform sampling
\begin{align}\label{eqn_noise_injected_values_130}
\diag(\bbc) \bbF \bbE_K = \sqrt{\frac{K}{N}} \bbF^{(K)},
\end{align}
where $\bbF^{(K)}$ is the Fourier basis of size $K \times K$. Hence, $\bbM = K/N \bbI$ [cf. \eqref{eqn_noise_injected_values_110}] and every eigenvalue of $\bbM$ equals $K/N$.
\end{myproof}

In words, even though for the noiseless case any seeding selection strategy satisfying the conditions in Proposition~\ref{P:Recov_seeds_setup1} is equally optimal, uniform seeding in directed cycles is the best MN-ST scheme when noise is present in $\underline{\bbs}_P$.

\begin{table}[t]
\centering
\begin{tabular}{l c c c }
\hline
& MN-ST & SN-MT & MN-MT \\\hline
\% of recovery & 91.8 & {\bf 96.4} & 94.4\\
Min error & {\bf .001} & .032  & .003 \\
Median error & {\bf .048} & .349 & .066 \\
\hline
\end{tabular}
\caption{Recovery performance for the three seeding schemes. {We restrict MN-MT to consist of two seeding nodes injecting two values.}}
\label{tab_recovery_comparison}
\vspace{-0.1in}
\end{table}

A second scenario of interest are setups where the additive noise at different value injections is uncorrelated and of fixed power, i.e., $\bbR_{\bbw} = \sigma^2 \bbI$. In this case, \eqref{eqn_noise_injected_values_030} can be rewritten as
\begin{equation}\label{eqn_noise_injected_values_040}
\bbR_{\bbeps} = \sigma^2 \bbV_K \boldsymbol{\Phi}_{K} \diag(\bbc) \boldsymbol{\Phi}^H _{K} \bbV^H_K.
\end{equation}
The design of $\bbc$ that minimizes the MSE of the reconstruction is the solution of the following linear integer program
\begin{align}\label{eqn_noise_injected_values_050}
\bbc^*& := \argmin_{\bbc} \,\, \trace(\bbR_{\bbeps}) = \argmin_{\bbc} \,\, \trace(\boldsymbol{\Phi}_{K} \diag(\bbc) \boldsymbol{\Phi}^H _{K}) \nonumber \\
&\text{s.t.} \qquad \bbc \in \{0, 1\}^{N\tau}, \quad \| \bbc \|_0 = P
\end{align}
which can be approximated by relaxing the binary and 0-norm constraints.

It turns out that the solution of \eqref{eqn_noise_injected_values_050} promotes the injection of seeding values at nodes that weakly express the active frequencies, i.e., nodes $j$ such that the values $[\widehat{\bbe}_j]_k$ for $k \leq K$ are small. This occurs because the noise power is fixed and those nodes require the injection of seeding signals with high power, leading to a high SNR.

\section{Numerical experiments}\label{S:NumExper}
We illustrate the reconstruction schemes in noiseless and noisy scenarios using synthetic (Section~\ref{sec_num_exp_synthetic}) and real-world graphs (Sections~\ref{sec_num_social_networks} and~\ref{sec_num_brain_state}).

\subsection{Synthetic graph signals}\label{sec_num_exp_synthetic}

Figure~\ref{fig:sub1_synthetic} represents a graph $\ccalG$ with $N=10$ nodes and adjacency matrix $\bbA$ generated using an Erd\H{o}s-R\'enyi (ER) model with edge probability 0.3 \cite{bollobas1998random}. Define the graph-shift operator $\bbS = \bbA$ and let $\bby$ be a signal to be recovered. Though seemingly random in the node domain, the structure of $\bby$ is highly determined by $\ccalG$. Indeed, $\bby$ has bandwidth $K=4$, as can be observed from its frequency representation $\widehat{\bby}$ in Figure~\ref{fig:sub1_synthetic}.

The first set of experiments illustrates the perfect recovery of $\bby$ when $P=4$ seeding values are injected into $\ccalG$ followed by a low-pass filter of degree $N-P=6$. The reconstruction is carried out using MN-MT seeding (Section~\ref{S:shift_space_extrapolation}) where nodes 1 and 2 act as seeding nodes and each of them injects a seeding value for time instants $t \in \{0,1\}$. After the seeding phase, a filter that successively annihilates the $N-K=6$ frequencies not active in $\bby$ is implemented [cf. \eqref{E:graph_filter_as_product}]. The evolutions of the reconstructed signal and its frequency representation are depicted in Figures~\ref{fig:sub2_synthetic} and~\ref{fig:sub3_synthetic}, respectively. Notice that perfect reconstruction is achieved since the last column in both figures coincide with $\bby$ and $\widehat{\bby}$. Figure~\ref{fig:sub2_synthetic} illustrates that the reconstructed signal is sparse during the seeding phase, consisting of the first two time instants. More specifically, for $t=0$ the signal attains nonzero values only for the seeding nodes [cf. \eqref{eqn_signal_y_time_t}] and for $t=1$ the signal remains zero for every node outside of the one-hop neighborhood of the seeding nodes. During the filtering phase -- times $t=2$ to $t=7$ -- signal values are successively exchanged between neighboring nodes in order to finally recover $\bby$ at time $t=7$. Figure~\ref{fig:sub3_synthetic} helps to understand the operation of the filtering phase. The signal $\bbx$ obtained after the seeding phase ($t=1$) contains every frequency not active in the desired signal $\bby$. Thus, in every successive time instant, one of these frequencies is annihilated. E.g., at time $t=2$ the frequency with index $i=5$ is eliminated and at $t=3$ the frequency $i=6$ is eliminated. In this way, at time $t=7$ every frequency not active in $\bby$ has been annihilated and perfect recovery is achieved.


\begin{figure}
\centering
\includegraphics[width=0.36\textwidth]{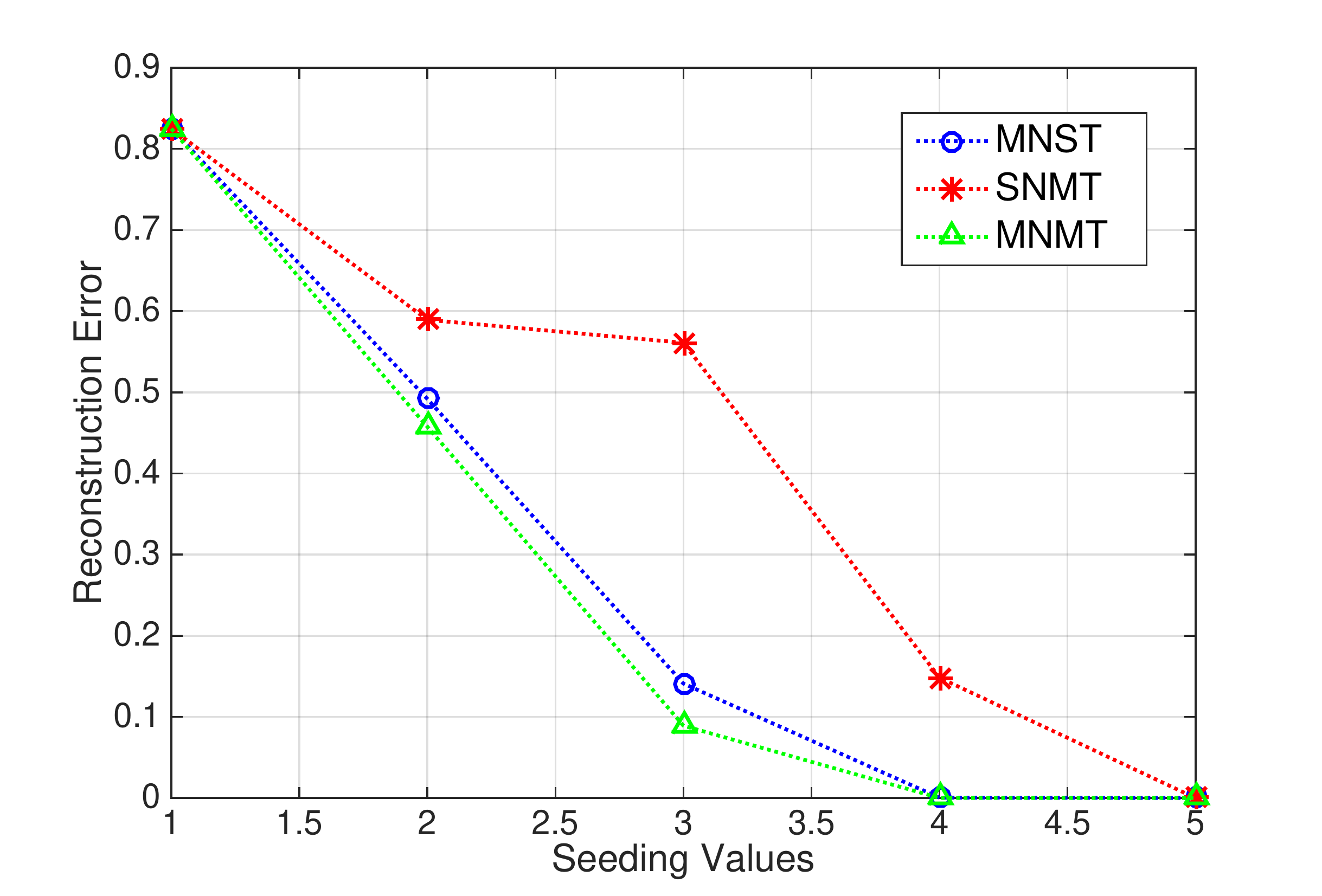}
\vspace{-0.05in}
\caption{Reconstruction errors when recovering a signal in a social network with insufficient seeding values.}
\label{fig_karate_reconstruction_error}
\vspace{-0.1in}
\end{figure}

To compare the reconstruction performance of MN-ST, SN-MT, and MN-MT seeding, we generate 1000 Erd\H{o}s-R\'enyi graphs with 10 nodes and edge probabilities between 0.2 and 0.4. On each graph we define a 4-bandlimited signal and try to recover it through the three seeding schemes presented; see Table~\ref{tab_recovery_comparison}. We restrict the MN-MT schemes to those consisting of two seeding nodes injecting two values each. We first compute the recovery percentage of the three schemes in noiseless scenarios. More specifically, for a given graph and signal to recover, we test for perfect recovery for every possible combination of seeding nodes. For example, there are 210 ways (10 choose 4) of selecting the seeding nodes in MN-ST while there are only 10 ways of selecting the single seeding node in SN-MT. If, e.g., 9 out of these 10 ways lead to perfect recovery, then the recovery percentage for SN-MT on that particular graph is 90\%. The values in Table~\ref{tab_recovery_comparison} correspond to the averages of these percentages across the 1000 graphs generated. Notice that the highest recovery percentage of SN-MT suggests that condition \emph{ii)} in Proposition~\ref{P:Recov_single_seeds_setup3} is more commonly satisfied in \emph{random} ER graphs than the respective conditions in Propositions~\ref{P:Recov_seeds_setup1} and~\ref{P:Recov_seeds_setup_general}. We then introduce noise in the injections following the constant SNR model in Section~\ref{sec_noise_when_injecting} for $\sigma = 10^{-3}$. Denoting by $\bbz$ the signal obtained from the reconstruction and by $\bby$ the desired signal, we define the reconstruction error as $\boldsymbol{\epsilon} = \| \bbz - \bby\|_2 / \| \bby \|_2$. For every given graph and signal $\bby$, we record the minimum and median $\boldsymbol{\epsilon}$ for every possible choice of seeding nodes within each reconstruction scheme. In Table~\ref{tab_recovery_comparison} we report the median of these values across the 1000 graphs generated. As it turns out, MN-ST is an order of magnitude more robust than SN-MT both in terms of minimum and median error. Finally, observe that MN-MT seeding presents an intermediate behavior both in terms of recovery percentage and reconstruction error.

\subsection{Influencing opinions in social networks}\label{sec_num_social_networks}

\begin{figure}[t]
\centering
\includegraphics[width=0.31\textwidth]{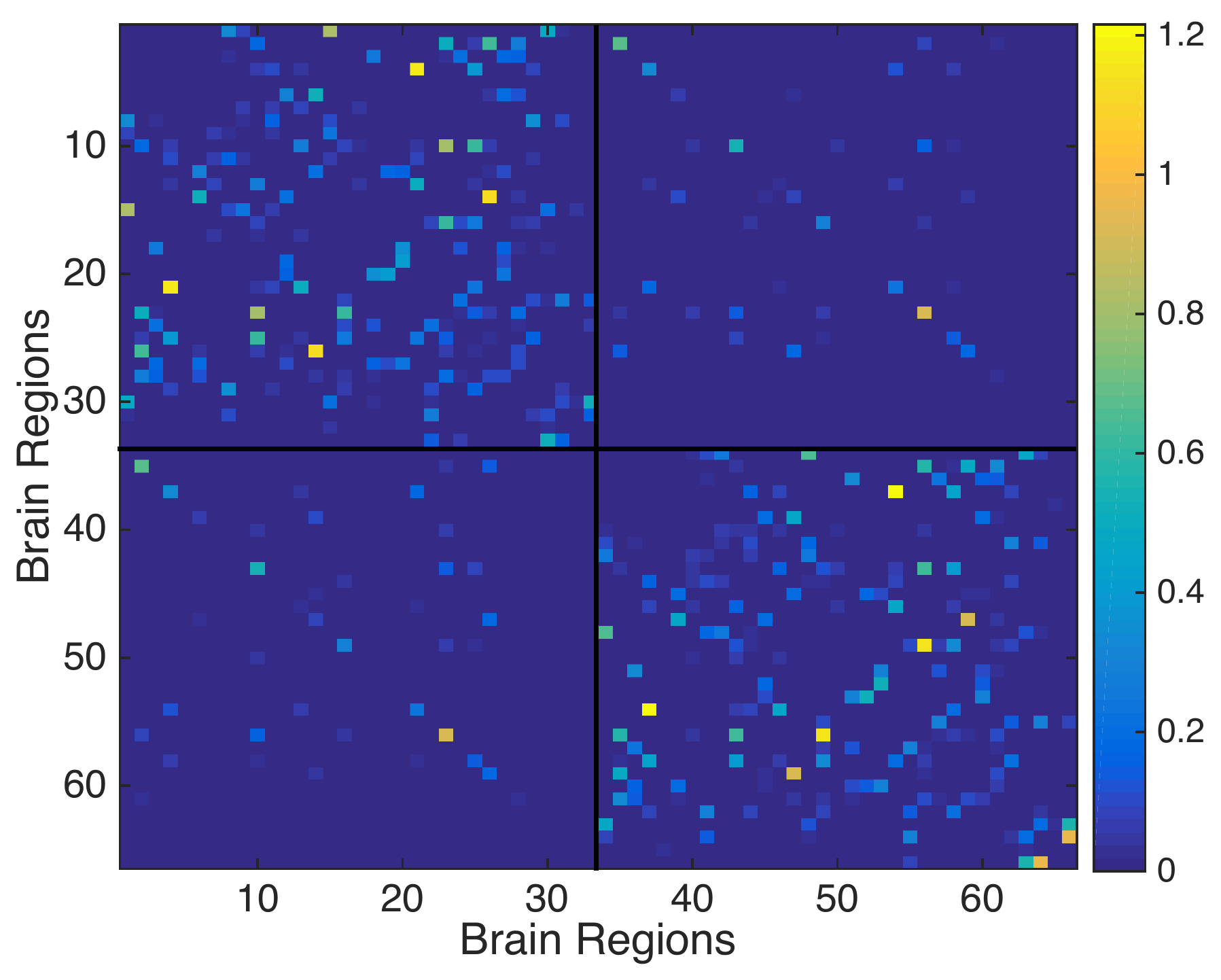}
\vspace{-0.05in}
\caption{Heat map of the adjacency matrix $\bbA$ of brain graph $\ccalG$.}
\label{fig_brain_network_imagesc}
\vspace{-0.1in}
\end{figure}

Consider the well-known social network of Zachary's karate club \cite{Zachary1977} represented by a graph $\ccalG$ consisting of 34 nodes or members of the club and 78 undirected edges symbolizing friendships among members. Denoting by $\bbL$ the Laplacian of $\ccalG$, define the graph shift operator $\bbS = \bbI - \alpha \bbL$ with $\alpha = 1/\lambda_{\max}(\bbL)$.  A signal $\bby$ on $\ccalG$ can be interpreted as a unidimensional opinion of each club member regarding a specific topic, and each successive application of $\bbS$ can be seen as an opinion update influenced by neighboring individuals. Bandlimitedness of $\bby$ implies that the opinion discrepancies between neighbors are small. In this context, signal reconstruction can be interpreted as the problem of inducing a desired global opinion profile by influencing the opinion of a subset of members.

\begin{figure}[t]
\centering
\includegraphics[width=0.46\textwidth, height = 0.2\textwidth]{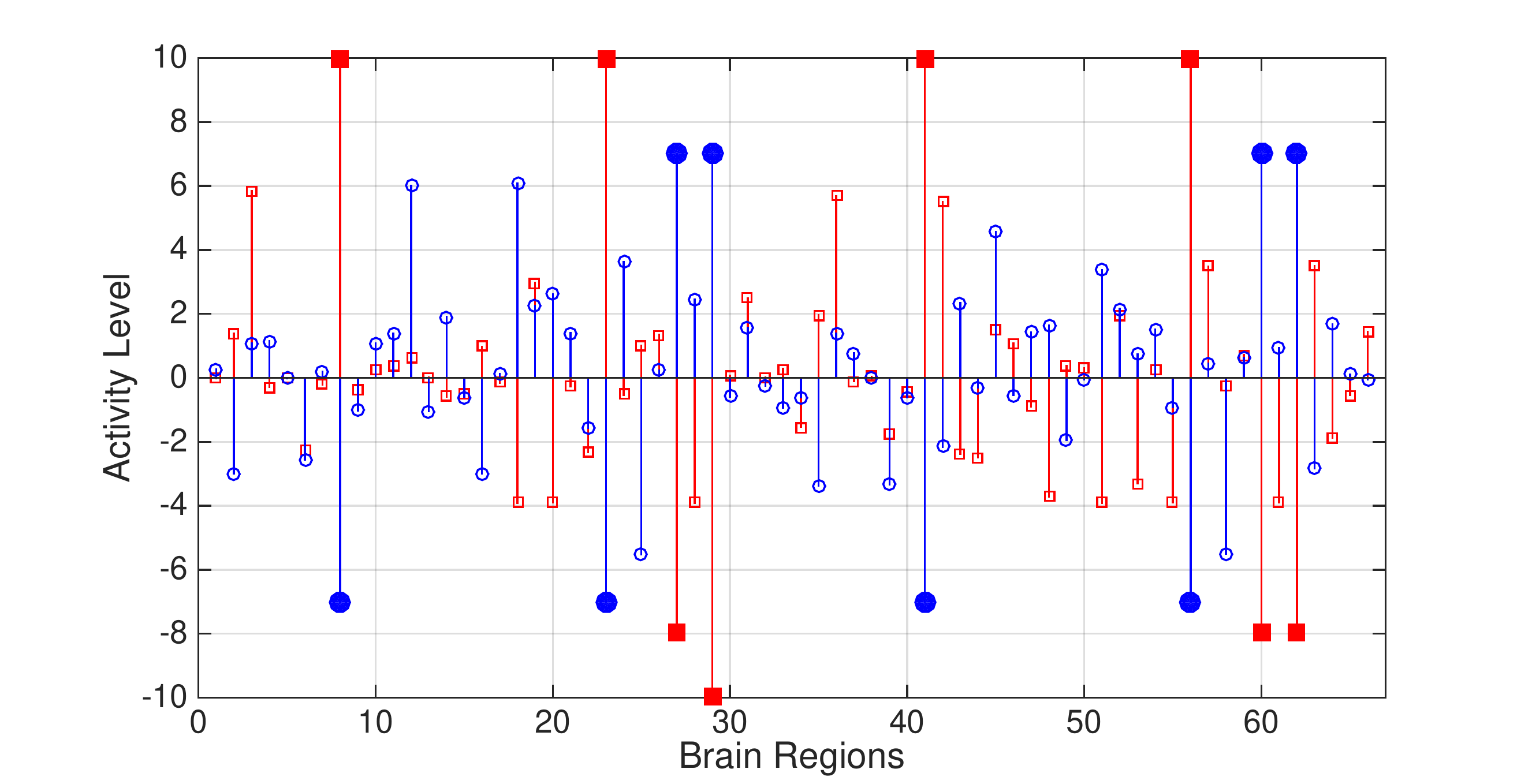}
\vspace{-0.05in}
\caption{Initial $\bby_i$ (red) and target $\bby_t$ (blue) brain states. High activity is represented by positive activity levels while negative values represent low levels of activity.}
\label{fig_initial_and_target_signals_brain}
\vspace{-0.15in}
\end{figure}

\begin{figure*}[t]
\centering

\begin{subfigure}{.3\textwidth}
  \centering
  \includegraphics[width=0.7\textwidth, height = 0.65\textwidth]{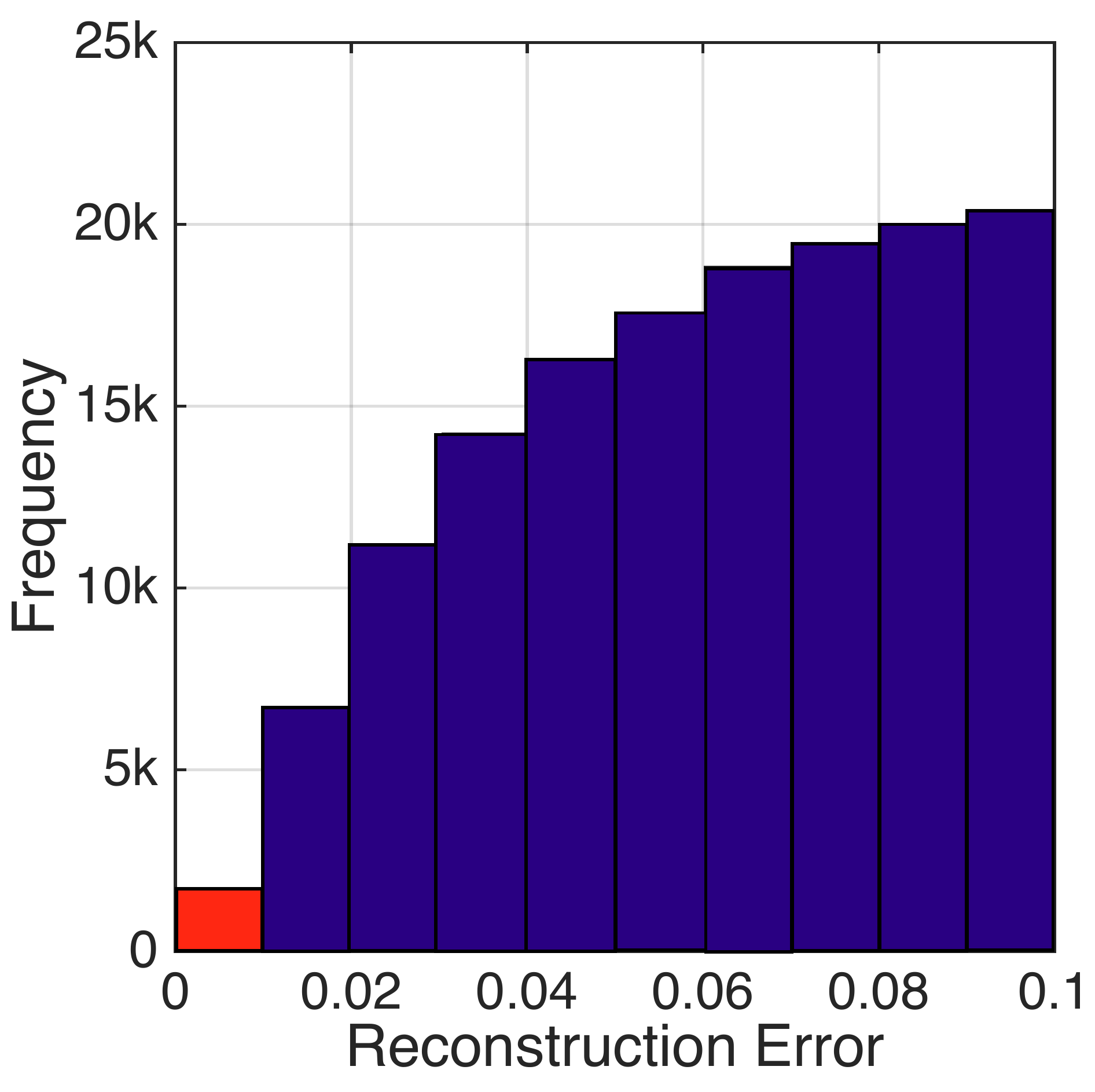}
  \caption{}
  \label{fig:sub1_brain}
\end{subfigure}%
\begin{subfigure}{.3\textwidth}
  \centering
  \includegraphics[width=1\textwidth, height = 0.65\textwidth]{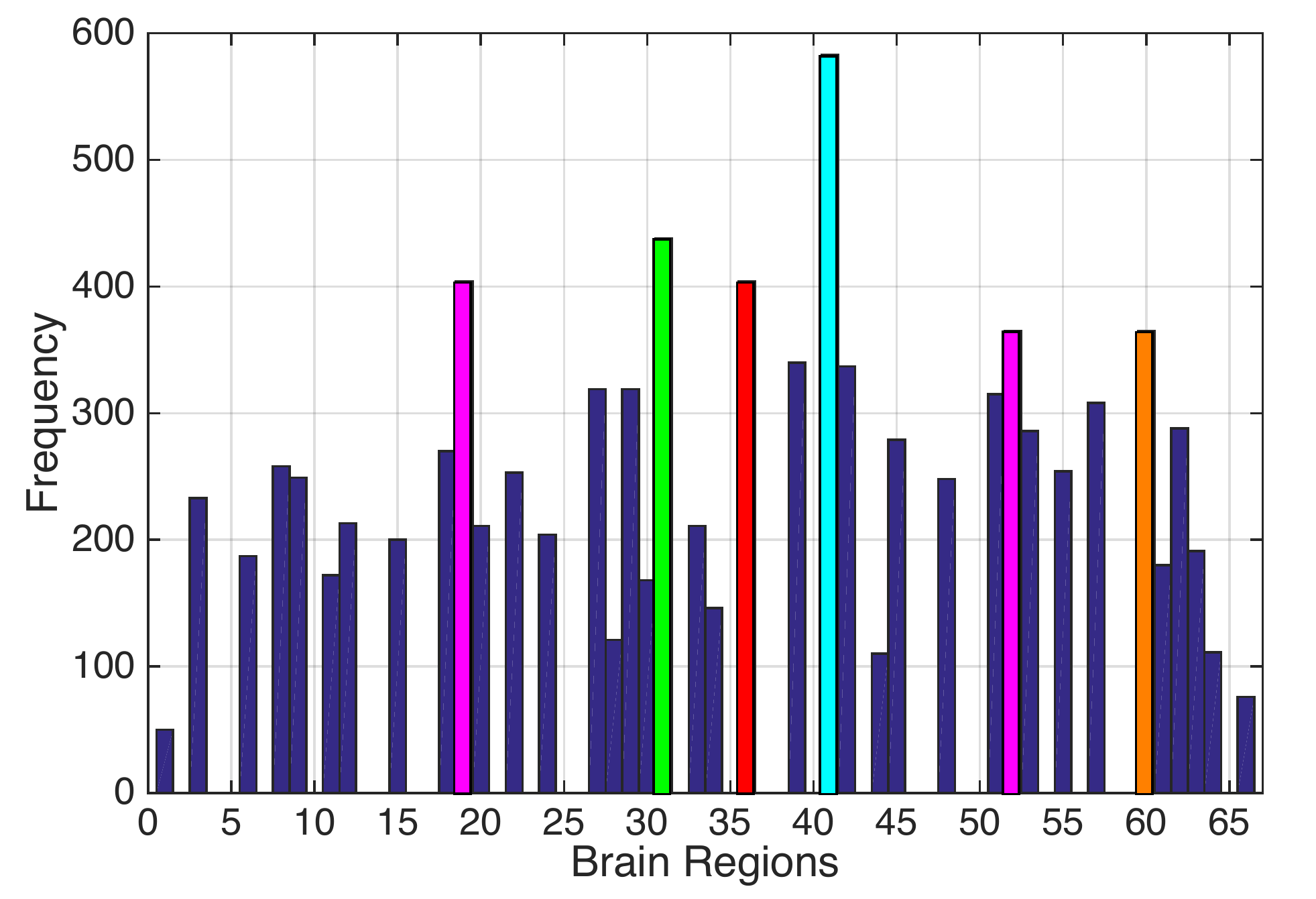}
  \caption{}
  \label{fig:sub2_brain}
\end{subfigure}%
\begin{subfigure}{.4\textwidth}
  \centering
    \input{figures/brain_regions_colored.tex}
  \caption{}
  \label{fig:sub3_brain}
\end{subfigure}
\vspace{-0.05in}
\caption{Inducing a brain state in the presence of noise. (a) Histogram of the reconstruction error for different choices of the six seeding nodes. (b) Frequency of appearance of each brain region among the configurations achieving the lowest reconstruction errors. (c) Anatomical location of the six regions most used in robust seeding configurations.}
\label{fig_brain_recovery}
\vspace{-0.15in}
\end{figure*}

We analyze the recovery performance when the number of seeding values is insufficient (Section~\ref{sec_insufficient_seeding_values}). For this, we generate a signal $\bby$ of bandwidth $K=5$ and try to reconstruct it using $P$ seeding values for $P = 1, \ldots, 5$; see Figure~\ref{fig_karate_reconstruction_error}. For every $P$, we find the combination of seeding values and locations that minimizes the reconstruction error within each seeding scheme (cf. Section~\ref{subsubsec_designing_seeding_values_seeding_locations}). E.g., if $P=2$ and we are analyzing SN-MT seeding, we consider every individual as possible seeding node and then choose the one achieving the minimum error. In Figure~\ref{fig_karate_reconstruction_error} we report the average of these minimum errors across 100 bandlimited signals. As expected, for $P=1$ the three schemes coincide and for $P=5$ perfect recovery is achieved for all of them. However, for intermediate values of $P$, MN-ST presents a considerably lower error than SN-MT. For $P=3$ this implies that, when trying to induce a global opinion profile, it is more effective to influence the opinion of three individuals once than to influence the opinion of the same individual three times. The fact that MN-MT seeding presents the lowest reconstruction errors is expected since this scheme includes the other two as particular cases. {Notice that in this case, as opposed to Table~\ref{tab_recovery_comparison}, we do not restrict MN-MT to the cases where multiple seeding nodes \emph{and} multiple seeding values in each node are used.}

For the above analysis to hold true, we must be able to apply a low-pass filter on the social network as required by the filtering phases of MN-ST, SN-MT, and MN-MT. This can be achieved by assuming that we can modify the rate of exchange of opinions in the network represented by $\alpha$. {Indeed, consider that after the seeding phase, the signal still percolates over the graph -- people still communicate their opinions to neighbors -- but we can modify the diffusion rate $\alpha_l$ at each discrete time instant. Thus, after $L-1$ interactions we obtain that
\begin{eqnarray}\label{eqn_graph_filter_social_network}
&\bbz = \prod_{l=1}^{L-1}(\bbI-\alpha_l \bbL) \bbx,&
\end{eqnarray}
which is equivalent to applying an annihilating filter [cf. \eqref{E:graph_filter_as_product}] to $\bbx$.} Notice that the filter in \eqref{eqn_graph_filter_social_network} is a polynomial on $\bbL$ rather than $\bbS$. However, the frequency annihilation procedure is still valid since the eigenvectors -- frequency basis -- of $\bbL$ and $\bbS$ are equal.

\subsection{Inducing a brain state}\label{sec_num_brain_state}

Upon dividing the human brain into the 66 regions of interest (ROIs) defined in \cite{hagmann2008mapping}, we build a weighted undirected graph $\ccalG$ whose nodes are the ROIs and whose edge weights are given by the density of anatomical connections between regions; see Figure~\ref{fig_brain_network_imagesc}. The first 33 ROIs are located on the right hemisphere of the brain while regions 34 to 66 correspond to their left counterparts. From Figure~\ref{fig_brain_network_imagesc} we see that most connections occur within the same cortical hemisphere with few inter hemispheric connections. We define the graph-shift operator $\bbS = \bbA$ where $\bbA$ is the adjacency matrix of $\ccalG$. The level of activity of each ROI can be represented by a graph signal $\bby$ where larger values represent higher levels of activity. Successive applications of $\bbS$ on $\bby$ model a linear evolution of the brain activity pattern \cite{gu2014controllability}. As a method to inject seeding values to $\ccalG$, we consider transcranial magnetic stimulation (TMS) \cite{hallett2000transcranial}, a noninvasive method to stimulate ROIs. In this context, reconstructing a brain signal amounts to inducing a specific brain state via TMS. In particular, we consider the problem of driving the brain from a resting state to one associated with high-level cognitive operations.

Brain resting states are associated with high activity in the posterior cingulate (PC) and inferior parietal (IP) cortices whereas active states are associated with high activity in the rostral middle frontal (RMF) and superior parietal (SP) cortices \cite{Greicius03, Medaglia15}.
In Figure~\ref{fig_initial_and_target_signals_brain} we present the initial $\bby_i$ and target $\bby_t$ signals, where the activity corresponding to the eight regions mentioned -- left and right versions of each cortex -- is highlighted with larger markers.
In order to drive the brain from $\bby_i$ to $\bby_t$ we consider a MN-MT seeding scheme with six seeding nodes. Since it is unclear how to implement a low-pass filter in a human brain, we consider that each seeding node injects eleven values, totalizing $P=66$ seeding values permitting the recovery of the target signal after the seeding phase without the need of a posterior filtering phase. Notice that throughout the paper we assumed the initial signal $\bby_i$ to be zero, meaning that there is no signal present on the graph before the reconstruction process. However, our model can accommodate for $\bby_i$ different from zero. To see this, if the seeding phase lasts $\tau$ instants, then we can design our seeding values to recover the signal $\bby_r = \bby_t - \bbS^{\tau-1} \bby_i$ in the original formulation so that the negative term cancels the effect of the seeding phase on $\bby_i$ and the target signal $\bby_t$ is recovered.

We consider noisy injections following the constant SNR model in Section~\ref{sec_noise_when_injecting} for $\sigma = 10^{-3}$. Denoting by $\bbz$ the reconstructed signal, define the reconstruction error as $\boldsymbol{\epsilon} = \| \bbz - \bby_t \|_2 / \| \bby_r \|_2$. We compute $\boldsymbol{\epsilon}$ for every possible combination of seeding nodes. Given that the seeding values are induced by TMS, we discard as possible seeding nodes the regions inaccessible by TMS like the ones located in the medial cortex and subcortical structures. After discarding inaccessible ROIs, the six seeding nodes can be chosen out of 38 possible ROIs, amounting to 2,760,681 possible configurations. In Figure~\ref{fig:sub1_brain} we present a histogram of the reconstruction error for different seeding configurations where we only show those attaining errors below 0.1. The red bar in this histogram corresponds to the 1,611 configurations that achieve the lowest reconstruction errors. In Figure~\ref{fig:sub2_brain} we present the frequency of appearance of each ROI in these 1,611 robust seeding configurations. The regions with zero appearances correspond to the ROIs inaccessible to TMS, however, among the accessible regions the frequency of appearance is not uniform. For example, the left inferior parietal cortex in position 41 appears 583 times whereas the right bank of the frontal pole in position 1 is only used 50 times. In Figure~\ref{fig:sub3_brain} we depict the 6 regions more commonly used in robust seeding configurations. Notice that both the left and right versions of the Pars Orbitalis are commonly used as seeding nodes, suggesting the importance of this region for robust brain state induction.

\section{Conclusions}\label{S:Conclusions}

A novel approach for the recovery of bandlimited graph signals -- that admit a sparse representation in the frequency domain -- was proposed.
The focus was not on estimating an unknown graph signal but rather on inducing a known bandlimited signal through minimal actions on the graph. These actions referred to signal injections at different seeding nodes, which then percolate through the graph via local interactions described by a graph filter. Restrictions on the number of seeding nodes and the amount of injections at each node gave rise to three different reconstruction schemes and their performance in noiseless and noisy settings was analyzed. For the noiseless case, we showed that a $K$-bandlimited signal can be recovered using $K$ injections followed by a low-pass filter in the (graph) frequency domain. In contrast to classical time-varying signals, it was also shown that if the seeding nodes inject the values of the original signal in those nodes, perfect recovery is not feasible. For scenarios leading to imperfect reconstruction, we analyzed robust seeding strategies to minimize distortion. Finally, the different reconstruction schemes were illustrated through numerical experiments in both synthetic and real-world graph signals.

\bibliographystyle{IEEEtran}
\bibliography{citations}

\end{document}